\documentclass[a4paper,fleqn,usenatbib]{mnras}

\usepackage[T1]{fontenc}
\usepackage{ae,aecompl}

\usepackage{longtable}
\usepackage{natbib}
\usepackage{psfrag}
\usepackage{xspace}
\usepackage{xcolor}
\usepackage{graphicx}   
\usepackage{amsmath}    
\usepackage{amssymb}    
\usepackage[normalem]{ulem}

\title[Stellar wind effects in young and old systems]{Atmospheric mass loss and stellar wind effects in young and old systems {I: comparative 3D study of TOI-942 and TOI-421 systems}}

\author[D. Kubyshkina et al.]{
Daria Kubyshkina$^{1,2}$\thanks{E-mail: kubyshkd@tcd.ie}, 
Aline A.~Vidotto$^{1,3}$, 
Carolina Villarreal D’Angelo$^{4}$, \newauthor \hspace{0.08 cm}
Stephen Carolan$^{1}$, 
Gopal Hazra$^{1,3}$,
Ilaria Carleo$^{5,6}$
\\
$^{1}$School of Physics, Trinity College Dublin, the University of Dublin, College Green, Dublin-2, Ireland\\
$^{2}$Space Research Institute, Austrian Academy of Sciences, Schmiedlstrasse 6, A-8042 Graz, Austria\\
$^{3}$Leiden Observatory, Leiden University, PO Box 9513, 2300 RA, Leiden, The Netherlands\\
$^{4}$ Instituto de Astronomía Teórica y Experimental (CONICET-UNC). Laprida 854, X5000BGR. Córdoba, Argentina\\
$^{5}$Astronomy Department and Van Vleck Observatory, Wesleyan University, Middletown, CT 06459, USA\\
$^{6}$ INAF – Osservatorio Astronomico di Padova, Vicolo dell’Osservatorio 5, I-35122, Padova, Italy\\
}

\date{Accepted XXX. Received YYY; in original form ZZZ}

\pubyear{2021}

\begin{document}
\label{firstpage}
\pagerange{\pageref{firstpage}--\pageref{lastpage}} \maketitle

\begin{abstract}
At young ages, {when radiation from the host star is high,} and the planet is hot and inflated after formation, planetary atmospheric mass loss can be extremely strong compared to {older planets.}
In turn, stellar winds are faster and denser for young stars compared to evolved main-sequence stars. Their interaction with escaping planetary atmospheres can substantially affect atmospheric mass loss rates, as well as the observable signatures of escaping atmospheres, with both effects expected to occur differently for young and evolved planets.  
We perform a comparative study of two systems around stars of similar masses but very different ages (50~Myr and 9~Gyr): TOI-942 and TOI-421. Both stars host two sub-Neptune-like planets at similar orbits and in similar mass ranges, which allows a direct comparison of the atmospheric escape and interactions with the stellar winds in the young and old systems.
We perform the 3D atmospheric modeling of the four planets in TOI-942 and TOI-421 systems and make the theoretical predictions of possible observational signatures in Ly-$\alpha$ absorption. We find that accounting for the stellar wind interacting with planetary atmospheres is crucial for the interpretation of the observations for young planets. Additionally, we show that a particular energy distribution along the XUV spectra has a minor effect on the atmospheric mass-loss rates, but it is of crucial importance for modeling the Ly-$\alpha$ absorption and therefore for interpretation of observations.
\end{abstract}




\section{Introduction}\label{sec::intro}

Atmospheric mass loss (when atmospheric material leaves the gravitation well of the planet and disperses into space), along with planetary formation processes and planetary system dynamics, are believed to be the key processes driving planetary atmospheric evolution and therefore shaping the observed population of exoplanets \citep[see, e.g.,][]{fulton2017,owen_wu2017,jin_mordasini2018,gupta_schlichting2019,gupta_schlichting2020,loyd2020,sandoval2021}. {Mass loss is particularly intense when the planet is young and hence hot and inflated (due to its post-formation luminosity) and is subject to the strong high-energy irradiation of the young host star \citep[e.g.,][]{watson1981,lammer2003,owen_wu2016,fossati2017,kubyshkina2018_k2-33b}.}

{At the same time, planets do not exist in empty space but are embedded into the extended astrospheres of their host stars. The escaping atmospheres interact with the stellar wind, which is the hot coronal material (low-density plasma) accelerated to high velocities carrying stellar magnetic field \citep[see, e.g.,][]{vidotto2015sw}.} For evolved main-sequence stars, such as the Sun, the speed of the stellar wind reaches a velocity of a few hundred ${\rm km~s^{-1}}$ at the distance of 1~AU \citep[see, e.g.,][]{gosling1976,zieger1998,abbo2016}. Young stars are characterized by higher coronal activity and higher mass loss rates \citep[][]{vidotto2021review}, and their winds, therefore, are expected to be denser, hotter, and faster compare to the Gyr-old stars. 

The interaction between stellar and planetary winds leads to changes in the planetary outflow's direction at large distances and to the appearance of a shock wave, whose position is controlled by the pressure balance. This can have a range of consequences for the planetary atmospheric dynamics. Thus, it was shown that stellar winds are capable of affecting the evaporation of the companion planets \citep[see, e.g.,][]{Matsakos2015,McCann2019,carolan2021}. Along with the possible erosion of planetary atmospheres by the stellar wind (increasing the mass loss rate), stellar winds can also suppress the escape in some cases. In particular, if the pressure balance is achieved close enough to the planet, where the planetary outflow remains subsonic, the stellar wind can significantly reduce atmospheric escape \citep{Christie2016,vidotto_cleary2020,carolan2020,carolan2021}. This can occur, for example, when the planetary outflow interacts with the strong wind of a young star \citep{carolan2020}, or if the planetary outflow is weak \citep{vidotto_cleary2020}.

{The stellar wind can also significantly affect the observational signatures of escaping atmospheres.}
The primordial atmospheres of planets, along with the atmospheres of the evolved planets in the sub-Neptune to Jupiter-like mass range, are mainly composed of hydrogen. Therefore, these escaping atmospheres can be probed using the spectroscopy of transiting planets in hydrogen lines, such as Ly-$\alpha$ or H$\alpha$. Absorption by atmospheric neutral hydrogen is expected to be more intense for young planets, due to their higher mass loss rates and hot expanded atmospheres \citep{allan_vidotto2019}. However, the actual observations can be complicated because of the high variability of young stars. Furthermore,  \citet{carolan2021} have demonstrated that due to the confinement of the atmosphere by strong stellar wind these signatures can be reduced, or even fully ``erased'', for planets in sub-Neptune/hot-Jupiter mass ranges. In more moderate cases, the absorption profiles become asymmetric with dominant blue-wing absorption (towards the observer) due to the atmosphere being pressed by the stellar wind in the direction {trailing the orbit} \citep[see, e.g.,][]{Carolina2018,Carolina2021}.

In this paper, we directly analyze the differences in the atmospheric escape and theoretical predictions of Ly-$\alpha$ absorption between planets around stars of similar properties and different ages. As such, we took the 9-Gyr old TOI-421 \citep{Carleo_toi421} and the 50-Myr old TOI-942 \citep{Carleo_toi942}. {The two stars have similar masses ($0.852^{+0.029}_{-0.021}$ and $0.88\pm0.04$~$M_{\odot}$) and close spectral types (G9 and K2.5V), allowing us to consider them as two evolutionary stages of the same star. Each hosts two sub-Neptune-like planets at nearly the same close-in orbits (about $0.05$ and $0.1$~AU) and in similar mass ranges (of about 7.2~$M_{\oplus}$ and 16.4~$M_{\oplus}$ in the older system, and below 16~$M_{\oplus}$ and 37~$M_{\oplus}$ in the younger system, for inner and outer planets, respectively). All four planets are expected to undergo strong atmospheric escape.}

We employ 3D models to study in detail the interaction between the evaporating planetary atmosphere and the stellar winds and make theoretical predictions for the observable Ly-$\alpha$ signatures. 
{We employ two different approaches to prescribe the stellar XUV (X-ray + extreme ultraviolet, EUV) irradiation: the commonly used single-wavelength model, which assumes the whole XUV flux is emitted with a photon energy of 20~eV, and the ``split-wavelength'' model, where the fluxes integrated over X-ray and EUV intervals are emitted with photon energies of 248~eV and 20~eV, respectively \citep[as was introduced in][]{kubyshkina2018grid}. The latter approach allows for more realistic energy redistribution and is expected to be relevant for young planets, where the X-ray part of the spectra holds a more significant fraction of the total energy \citep[see, e.g.,][]{sanz2011,matt2015,Johnstone2020mors}. Here we confirm this and show that though the effect of the particular energy distribution along the XUV spectra has a minor effect on the atmospheric mass-loss rates, it is of crucial importance for modeling the Ly-$\alpha$ absorption and therefore for interpretation of observations.}

The paper is organized as follows. In Section~\ref{sec::target_systems} we discuss the properties of the stellar hosts TOI-421 and TOI-942 and their planets, and in Section~\ref{sec::model} we describe the models used in this study. We present our results and discuss their possible implications in Section~\ref{sec::results} and summarize our conclusions in Section~\ref{sec::conclusions}.

\section{Target systems}\label{sec::target_systems}

\subsection{Stars}\label{sec::target_systems_star}
The two systems, TOI-421 and TOI-942, despite formally belonging to different stellar types (G9 and K2.5V, respectively), show a large similarity if accounting for their ages {(see parameters in Table~\ref{tab::systems})}. The masses ($M_*$) and radii ($R_*$) of these stars coincide within the uncertainties, and the difference in the effective temperatures is about 350~K, with both temperatures being close to the border value of 5000~K between G and K classes. The latter, as well as the bolometric luminosities of these stars, are typical for a star of $\sim 0.85 M_{\odot}$ at the given ages of TOI-421 and TOI-942, which are of $9.4^{+2.4}_{-3.1}$~Gyr and $50^{+30}_{-20}$~Myr, respectively, as given by stellar evolution models \citep[see, e.g.,][]{spada2013,choi2016}. Here, we employ the parameters of TOI-421 and TOI-942 systems from \citet{Carleo_toi421} and \citet{Carleo_toi942}, respectively, unless explicitly specified otherwise\footnote{\citet{zhou2021toi942} suggest the same rotation period, but the lower stellar mass and the larger stellar radius for TOI-942 (both differ by about 10\% from the values used here), and consequently a slightly larger radii of the planets in the system. They, however, provide only a weak constraint on the planetary masses of $\sim$2.5~$M_{\rm jupiter}$ upper limit. For this reason, and to have consistent data for both TOI-942 and TOI-421 system (which is guaranteed by the same methods used in \citet{Carleo_toi421} and \citet{Carleo_toi942}), we employ the data from \citet{Carleo_toi942} for TOI-942 planetary system.}. 

\begin{table*}
\caption{Parameters of TOI-421 and TOI-942 systems as given in
\citet{Carleo_toi421} and \citet{Carleo_toi942}, respectively.}
\begin{tabular}{|c|c|c|}
  \hline
                                 & TOI-942                   & TOI-421 \\
  \hline
  $\rm M_*$ $\rm [M_{\odot}]$    & $0.88\pm 0.04$            & $0.852^{+0.029}_{-0.021}$ \\
  $\rm R_*$ $\rm [R_{\odot}]$    & $0.893^{+0.071}_{-0.053}$ & $0.871\pm0.012$ \\
  $\rm T_{eff}$ [K]              & $4969\pm100$              & $5325^{+78}_{-58}$ \\
  $\rm L_{bol}$ $\rm[L_{\odot}]$ & $0.4375^{+0.0357}_{-0.0206}$ & $0.5339^{+0.0157}_{-0.0098}$ \\
  age [Gyr]                      & $0.05^{+0.03}_{-0.02}$    & $9.4^{+2.4}_{-3.1}$ \\
  $\rm P_{rot}$ [days]           & $3.39\pm0.01$             & $43.24^{+0.57}_{-0.55}$ \\
  $\rm log_{10}(L_X [{\rm erg~s^{-1}}])$    & 30.07                     & 27.9 \\
                                 &                           &     \\
  PLANET                         & b\hspace{1.5 cm} c                     & b\hspace{1.5 cm} c \\
  $\rm M_{pl}$ $[M_{\oplus}]$     & $<16$\hspace{1.5 cm} $<37$             & $7.17\pm0.66$\hspace{1.5 cm} $16.42^{+1.06}_{-1.04}$ \\
  $\rm R_{pl}$ $[R_{\oplus}]$     & $4.242^{+0.376}_{-0.313}$\hspace{1.5 cm} $4.793^{+0.410}_{-0.351}$ & $2.68^{+0.19}_{-0.18}$\hspace{1.5 cm} $5.09^{+0.16}_{0.15}$ \\
  $\rm a$ [AU]                 & $0.0498\pm0.0007$ \hspace{1.4 cm} $0.088\pm0.0014$ & $0.056\pm0.0018$\hspace{1.4 cm} $0.1189\pm0.0039$ \\
  $\rm F_{XUV}$ [${\rm erg~s^{-1}cm^{-2}}$]                 & $2.7\times10^5$ \hspace{1.4 cm} $8.5\times10^4$ & $7452$\hspace{1.4 cm} $1654.8$ \\
  \hline
\end{tabular}
\label{tab::systems}
\end{table*}

Given the above, we employ the stellar {rotation (activity)} evolutionary models by \citet[][Mors code]{Johnstone2020mors,spada2013}
to {search for} evolutionary track reproducing parameters of both systems.
{These models are constrained by the observed distribution of rotation periods in young stellar clusters, and the set of empirical relations based on X-ray, EUV, and Ly-$\alpha$ measurements.}

{The rotation period of the star and its high-energy radiation (X-ray, EUV) are closely connected \citep[see, e.g.,][]{pizzolato2003, mamajek2008, wright2011,jackson2012,tu2015, magaudda2020,Johnstone2020mors}. When {single} stars are born, they have high rotation rates, strong magnetic fields, and high XUV luminosity, which all decay with time. 
{The decay in X-ray luminosity $L_{\rm X}$ can be approximated by a power law as $\frac{L_{\rm X}}{L_{\rm bol}}\sim(\frac{P_{\rm rot}}{P_{\rm conv}})^{\beta}$, where $P_{\rm conv}(M_*)$ is the convective turnover time and $\beta$ is a mass-dependent constant. Note that $\beta$}
differs for the initial saturation period (when stellar $L_{\rm X}$ depends weakly on rotation) and for later times. Thus, the stellar high-energy radiation at a specific time depends on stellar mass and rotation. For young stars, it also depends strongly on the initial $P_{\rm rot}$ of the star, which leads to a wide spread in stellar XUV luminosities at early ages. After $\sim$1~Gyr, this spread is substantially reduced and stars born with different $P_{\rm rot}$ converge to a tight rotation-age relation \citep[e.g.,][]{matt2015,Johnstone2020mors}.
 }

We start by fitting the rotation period of the younger system TOI-942. As both stars have similar masses and TOI-421 rotation period should be already converged at its age, fitting the period of the younger will automatically fit the period of the older \citep[see, e.g.,][]{tu2015,Johnstone2020mors}.
{However, }fitting only $P_{\rm{ rot}}(age)$ for TOI-942 does not fit properly other stellar parameters important for our planetary evolution models, as the Mors code is based on large stellar samples and predicts therefore {an average value of the parameters}. To solve this issue we vary the stellar mass and age within the observational uncertainties for TOI-942. This has a minor effect on the predicted $P_{\rm{ rot}}(age)$.

{We aim to fit the parameters that have a larger impact on the escape rate as well as possible. This includes the effective temperature/bolometric luminosity of the star, which are responsible for the equilibrium temperature of a planet, and the XUV luminosity, which depends on both $L_{\rm bol}$ and $P_{\rm rot}$, and significantly affects the atmospheric escape. }

\begin{figure*}
  \includegraphics[width=0.8\hsize]{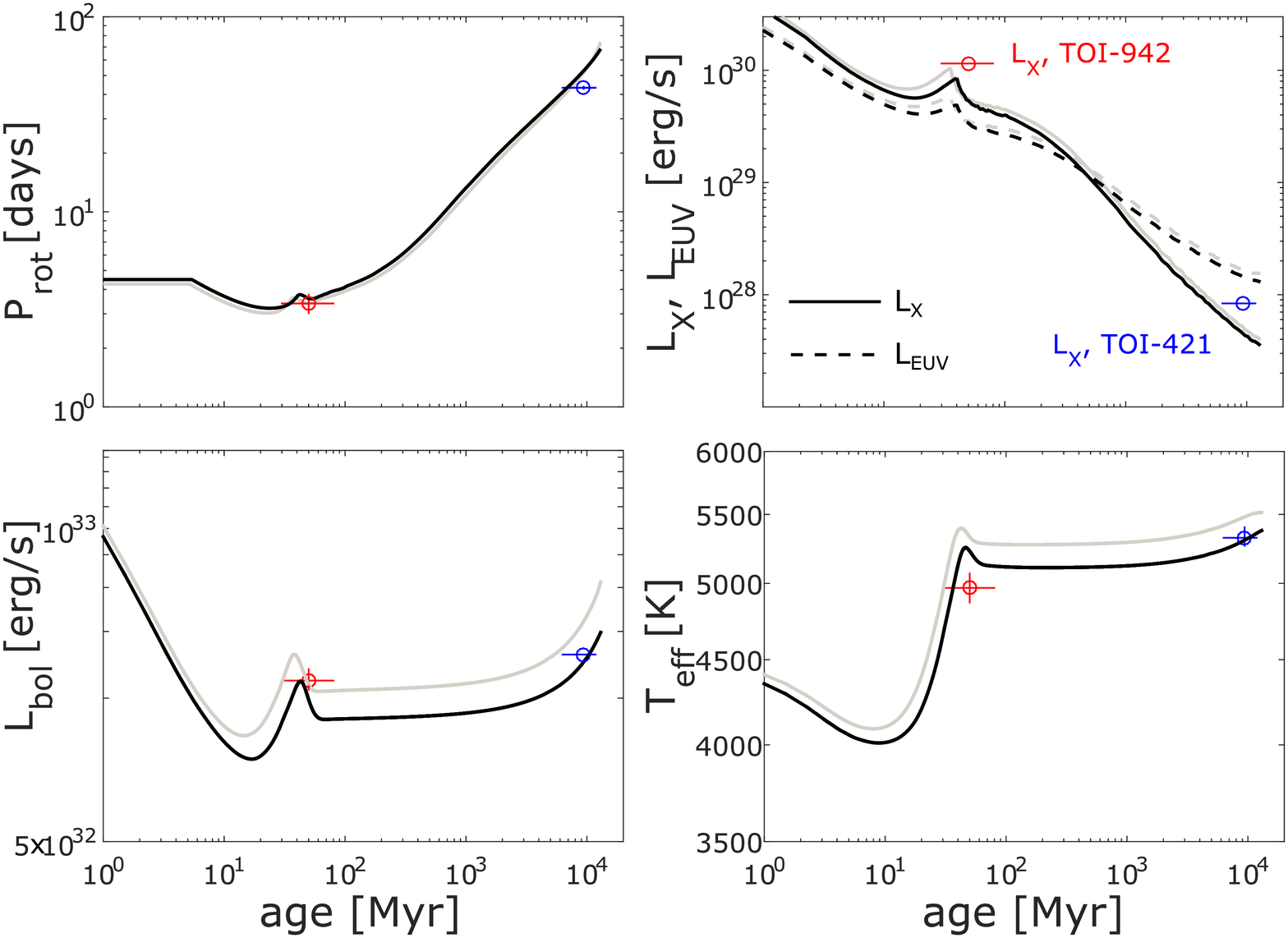}\\
\caption{Evolutionary track of TOI-421 \citep{Carleo_toi421} and TOI-942 \citep{Carleo_toi942} systems fitted using the Mors stellar evolution code \citep{Johnstone2020mors}. In the top two panels we show the {evolution of the} rotation period and X-ray/EUV luminosities, and in the bottom panels, the {evolution of the} bolometric luminosity and the effective temperature of the star. 
{The black lines show the model corresponding to the stellar mass of 0.84~$M_{\odot}$, and a stellar rotation period of 3.39 days at the age of 35~Myr. For comparison, we show also the model (pale gray lines) corresponding to the best-fit parameters of TOI-942 alone listed in Table~\ref{tab::systems}.}
The red/blue circles show the present values for TOI-942/TOI-421 stars, as indicated in the top right panel. The error bars for $P_{\rm rot}$, $T_{\rm eff}$, $L_{\rm bol}$, and the ages of the two systems are shown with horizontal/vertical lines. The given values of $L_{\rm X}$ are model dependent and are given for the median values.}\label{fig::stellar_models}
\end{figure*}

In Figure~\ref{fig::stellar_models}, we show in black the track (stellar parameters against time) which best reproduces the parameters of TOI-942 and TOI-421 at their ages. {This track} corresponds to the assumption that the mass of TOI-942 is at its lower uncertainty border equal to 0.84~$M_{\odot}$, and the age of the star is 35~Myr (``track 1''). For comparison, we also show the model that corresponds to the best-fit parameters of TOI-942 (``track 0'', in pale gray). {Even though the rotation periods of the two stars are well reproduced by both models, and the predictions for $L_{\rm X}$ and $L_{\rm EUV}$ are nearly the same in both tracks,
the bolometric luminosity and the effective temperature of a star, are better match by ``track 1'' for both stars. }

It is interesting to note that for both stars the X-ray luminosity predicted by the evolutionary models for the given rotation periods is lower than the $L_{\rm X}$ estimated from observations by a factor of $\sim$2. This systematic difference is likely coming from two sources. First, the relations between the rotation and X-ray luminosity in models by \citet{Johnstone2020mors} were fit on a large sample of stars, and represent therefore an averaged relation. Second, the estimates based on the observations remain model dependent. In this context, we find an agreement of a factor of two adequate for the two systems. In the absolute values, the X-ray luminosity of the young star TOI-942 is $\sim575$ times higher than that of TOI-421.
{We do not show the values of $L_{\rm EUV}$ in Figure~\ref{fig::stellar_models}, as no direct measurements of EUV are available, and we rely on the empirical relations presented in \citet{Johnstone2020mors} to derive their values. We note that different approximations \citep[see also, e.g.,][]{sanz2011,linsky2014,linsky2020} can result in $L_{\rm EUV}$ values that differ by a factor of {$\sim2-4$ (particularly at early ages)}. This difference, however, does not affect the conclusions of the present study.}

\begin{figure}
  \includegraphics[width=\hsize]{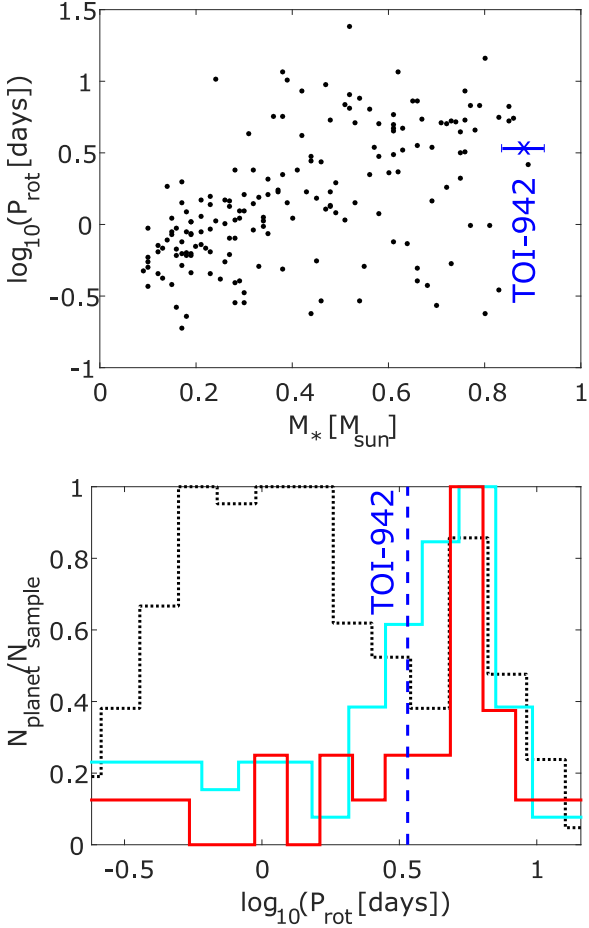}\\
  \caption{Distribution of stellar rotation periods in the young cluster NGC 2547 {of similar age to TOI-942}. Top panel: rotation period (in logarithmic scale) against stellar masses in the sample. The blue bar represents TOI-942. Bottom panel:  histogram of relative occurrence of stars rotating with a specific period for the whole sample (black dotted line), and for cluster members heavier than 0.5~$M_{\odot}$ (cyan) and than 0.7~$M_{\odot}$ (red). The period of TOI-942 is marked by the vertical blue dashed line. }\label{fig::irwin2008}
\end{figure}

To put the young star TOI-942 in the context of stars of similar ages, we compare it to the observations of the young cluster NGC 2547 \citep[][]{irwin2008}. The sample consists of 176 stars in a mass range of 0.1-0.9~$M_{\odot}$ at the age of $\sim 40$~Myr. We show the distribution of the rotation periods against mass and the normalized histograms for different mass ranges in Figure~\ref{fig::irwin2008}. Comparing to stars of similar age and mass, TOI-942 with a rotation period of $\sim3.4$~days is faster than average, but not extreme (the shortest period in the sample for $M_*>0.7M_{\odot}$ is 0.24 days). TOI-421 has, in turn, the typical rotation period for its age, when the spread in rotation periods for stars of a specific mass is small (see, e.g., \citealt{matt2015,Johnstone2020mors}).

\subsection{Planets}

Both TOI-942 and TOI-421 host two sub-Neptune-like planets in similar close-in orbits (planets ``b'' at $\sim$0.05 and planets ``c'' $\sim$0.1~AU). The detailed parameters of the planets are given in Table~\ref{tab::systems}. For the younger system TOI-942, only the upper limits of planetary masses are constrained. In this study, we 
{assume their} masses are similar to the masses of planets in the evolved TOI-421 system to facilitate comparison, namely 8.5 and 17~$M_{\oplus}$ for planets b and c, respectively. These masses are about half of the upper limit, and the choice of these particular values has a small effect on overall results and conclusions, unless we consider atypically low planetary densities. For example, adopting the mass of TOI-942~b at the upper conservative limit of 16~$M_{\oplus}$ (instead of 8.5~$M_{\oplus}$) reduces the atmospheric escape rate from the planet by about 11~\%. 

The radii of {the} two outer planets coincide within the uncertainty. Taking into account that planetary radii depend mostly on the atmospheric mass fraction of a planet {($f_{\rm at} = \frac{M_{\rm atm}}{M_{\rm pl}}$, where $M_{\rm atm}$ is the mass of the atmosphere)} and its temperature \citep[see, e.g.,][]{lopez2014,Chen_rog2016,kubyshkina2020mesa,kubyshkina2021mesa}, we suggest that both planets host a similar amount of the atmosphere: {$f_{\rm at} \sim $ 0.02-0.075 for TOI-942~c and $f_{\rm at} \sim$ 0.03-0.105} for TOI-421~c based on the approximation introduced in \citet{kubyshkina2021mesa}. 
Overall, the results presented in this paper predict that the planet with parameters similar to TOI-421~c could lose $\sim$40-80$\%$ of its initial atmosphere (with the particular value depending on the initial atmospheric mass fraction and the rotational evolution of its host star). Therefore, it is unlikely that TOI-942~c could actually be the ``past self'' of TOI-421~c. The same holds for the inner planets TOI-942~b ($f_{\rm at}$ of 0.01-0.05) and TOI-421b ($f_{\rm at}$ up to $0.01$), where the atmosphere would most likely be fully evaporated, given that TOI-942 rotates faster (and therefore emits more high-energy radiation) than average. We will, however, further examine this possibility with a more precise approach in Paper II of this series. 

{Finally, we should note that the planetary orbits in {the} two systems are similar but not exactly the same, with orbital distances of the outer planets differing by $\sim 0.03$~AU. Changing the positions of {the} planets in one system relative to those in another could result in changes in the incident irradiation ($F_{\rm XUV}$ in Table~\ref{tab::systems}) by about $20\%$ for the inner planets and by about $50\%$ for the outer planets. {This is negligible compared to the two-orders of magnitude difference in  $L_{\rm XUV}$ of both stars, and thus does not affect significantly the direct comparison we do between these planets. Furthermore, according to hydrodynamic models, the corresponding changes in mass-loss rates caused by the 20-\% and 50-\% changes in $F_{\rm XUV}$ would be of $\sim 9\%$ for the inner and $\sim 20\%$ for the outer planets. This is not substantial to affect the conclusions of this study.}
} 


\section{Model}\label{sec::model}

\subsection{Atmospheric model}\label{ssec::model-Aline}

{In this paper, we follow the modeling approach introduced in \citet{hazra2021} and \citet{carolan2021r}. We  point the reader to these papers for a more thorough description of the model and only briefly describe it here.}
The modeling is based on employing the user module of BATS-R-US \citep[][]{toth2005} to perform the 3D hydrodynamics simulation of the escaping upper planetary atmosphere and its interaction with the stellar wind, with parameters of the latter, as well as the boundary conditions, set by external models.

The basic physical model assumes a pure hydrogen atmosphere and is based on solving the mass, momentum, and energy conservation equations in the following form
\begin{eqnarray}\noindent
   &&\frac{\partial\rho}{\partial t} + \bigtriangledown(\rho\vec{u}) = 0 \label{eq::mass_cons}\\
   &&\frac{\partial(\rho\vec{u})}{\partial t} + \bigtriangledown[\rho\vec{u}\vec{u}+PI] = \nonumber\\ &&= \rho\vec{g} - \frac{\rho GM_*}{(r-a)^2}\hat{R} - \rho\vec{\Omega}\times(\vec{\Omega}\times\vec{R}) - 2\rho(\vec{\Omega}\times\vec{u})\label{eq::moment_cons}\\
   &&\frac{\partial E}{\partial t} + \bigtriangledown[\vec{u}(E+P)] = \nonumber\\ &&= \rho\left(\vec{g} - \frac{GM_*}{(r-a)^2}\hat{R}  \vec{\Omega}\times(\vec{\Omega}\times\vec{R})\right)\vec{u} + {\cal H} - {\cal C}.\label{eq::energy_cons}
\end{eqnarray}
Here $\rho$, $\vec{u}$, and $P$ are mass density, bulk velocity, and the thermal pressure. The momentum Equation~\ref{eq::moment_cons} accounts for planetary and stellar gravity, centrifugal and Coriolis forces (terms on the right-hand side of the equation, respectively), {where $\vec{g} = \frac{GM_{pl}}{r^2}\hat{r}$ is the term describing central gravity of the planet}, $G$ is the gravitation constant, $M_*$ is the stellar mass, and $\vec{\Omega}$ is the orbital rotation rate. $\vec{r}$ and $\vec{R}$ are the positions relative to the planet and star centers, respectively, and $a$ is the orbital distance.

The energy density in Equation~\ref{eq::energy_cons} is $E = \rho u^2/2 + P/(\gamma-1)$. The adiabatic index $\gamma$ is taken as $5/3$. To distinguish between stellar and planetary material we use a passive scalar.  In both cases, we assume  ideal gas and therefore {the} thermal pressure is $P = \rho k_{\rm B}T/(\mu m_{\rm p})$, where
$k_{\rm B}$ is the Boltzmann constant, $\mu$ is the mean mass per
particle and $m_{\rm p}$ is the mass of the proton.

{The right-hand side of Equation~\ref{eq::energy_cons} includes heating and cooling processes considered in the model: ${\cal H} - {\cal C}$. Here, ${\cal H}$ is the rate of photo-heating by stellar XUV irradiation. In this paper, we either consider the XUV irradiation is concentrated at a single-wavelength  \citep[which is a common approach used, e.g., in ][]{MC2009,allan_vidotto2019} or use a ``split-wavelength'' approach \citep[see Section 4.3 and][]{kubyshkina2018_k2-33b,kubyshkina2018grid}.}
Assuming the integrated XUV flux to be emitted at the single wavelength corresponding to a photon energy of $E_{\nu_0} = 20$~eV, we have that
\begin{equation}\label{eq::heating}
    {\cal H} = \eta F_{\rm XUV}e^{-\tau}\sigma_{\nu_0}n_{\rm n}, 
\end{equation}

\noindent where $F_{\rm XUV}$ is the XUV flux at the planetary orbit, $\tau$ is the optical depth for the aforementioned photon energy, and $n_{\rm n}$ is the number density of  neutral hydrogen. The photoionisation cross section for atomic hydrogen is $\sigma_{\nu_0} = 6\times10^{-18}(\frac{E_{\nu_0}}{13.6 {\rm eV}})^{-3}$~${\rm cm^2}$. $\eta$ is so-called heating efficiency, i.e., the fraction of the absorbed XUV energy spent to heat the atmosphere. Here, we follow the approach of \citet{MC2009} and adopt 
\begin{equation}\label{eq::eta}
    \eta = \frac{E_{\nu_0}-13.6 {\rm eV}}{E_{\nu_0}},
\end{equation}

\noindent which equals to $\sim 0.32$ for the $E_{\nu_0} = 20$~eV.

%
%
%

{The cooling rate ${\cal C}$ accounts for radiative losses due to the collisional excitation (Ly-$\alpha$)
\begin{equation}
    {\cal C}_{\rm Ly\alpha} = 7.5\times10^{-19}n_{\rm p}n_{\rm n}\exp[-1.183\times10^5/T] \hspace{0.1 cm}\mathrm{erg~cm^{-3}s^{-1}},
\end{equation}
and collisional ionisation 
\begin{equation}
{\cal C}_{\rm col.ion.} = 1.27\times10^{-21}n_{\rm p}n_{\rm n}\exp[-157809/T] \hspace{0.1 cm}\mathrm{erg~cm^{-3}s^{-1}}
\end{equation}
\citep{black1981}, where $n_{\rm p}$ is the number density of protons (${\rm H^+}$) and $T$ is the atmospheric temperature. }

In parallel with Equations~\ref{eq::mass_cons}, \ref{eq::moment_cons}, and \ref{eq::energy_cons}, the ionisation/recombination balance is solved including photo- and collisional ionisation at rates given in \citet{black1981}.

In all cases considered in this paper, we use the rectangular 3D grid with the planet located in the origin and the relative size of $[-50, 50]R_{\rm pl}$ in X and Y directions, and $[-32, 32]R_{\rm pl}$ in the Z direction. 
The coordinate system is equivalent to the Geocentric Solar Ecliptic (GSE) system with an inverted X-axis: the orbital plane is {the} XY plane with X-axis pointing opposite to the star, and Z-axis perpendicular to this plane, aligned with the planet spin axis. {The grid consists of $\sim$16 million cells with non-uniform resolution \citep[][]{carolan2020}. The minimum resolution in the grid is 2~$R_{\rm pl}$ in the outer zone, and it increases to 1/2~$R_{\rm pl}$ within the box of [-10, 20]~$R_{\rm pl}$ in the X direction, and [-10, 10]~$R_{\rm pl}$ in Y and Z directions, and further to 1/8~$R_{\rm pl}$ within the sphere of 10~$R_{\rm pl}$ radius, and to 1/16~$R_{\rm pl}$ within the 5~$R_{\rm pl}$ sphere.}

\subsection{Stellar winds}\label{ssec::model-SW}

Here we follow the approach used in \citet{carolan2020} to include the stellar wind in the 3D simulation: we adopt a stellar mass loss and coronal temperature and use 1D Parker wind model to estimate the wind parameters at the planetary orbit, which are further used to inject the wind into the simulation domain (i.e., to set up the outer boundary conditions).

To model the stellar wind from the more evolved star TOI-421, we employ the stellar parameters previously used in \citet{Carleo_toi421}, close to solar values. Thus, stellar mass loss is taken of $\dot{M}_* = 2\times10^{-14}M_{\odot}{\rm yr}^{-1}$ and the coronal temperature is $T_{\rm c} = 1.3$~MK. This results in the stellar wind accelerated to a supersonic velocity of $\sim$100~${\rm km~s^{-1}}$ near the orbits of b and c planets. 

In the case of the younger star, both the mass loss and temperature (and, hence, velocity) of the wind are higher. From the diagram given in \citet{vidotto2021review} relating the X-ray flux of the star and its escape rate (this diagram is a compilation of a range of observational works cited therein), we can expect that the mass loss from TOI-942 is about $10^{-12}M_{\odot}{\rm yr}^{-1}$, i.e., $\sim 50$ times higher than at TOI-421. The temperature of the stellar coronae (and, thus, wind, as it is assumed to be isothermal) remains unknown and set, therefore, more or less arbitrarily. To exclude the possible effect of this choice, we have considered two different temperatures of 3~MK (which is around the upper limit of solar coronal temperature at the active periods) and 4~MK. We found, however, that the choice of a particular temperature does not affect {the} results significantly, and in the following, we will only show the 4~MK case.

{The original 1D model of the stellar wind that we use is isothermal and assumes $\gamma = 1$. However, 
{our model}
does not allow the implementation of  different adiabatic indexes within the same modeling domain. Thus, in the 3D simulation, the stellar wind flow is modeled with $\gamma = 5/3$. This results in the stellar wind temperature slowly decreasing from the boundary where the wind is injected to the position of the planet, as will be shown in Section~\ref{ssec::results_sw}}.

\section{Results}\label{sec::results}

\subsection{Overview of the results}\label{ssec::results_overview}

\begin{figure*}
  \includegraphics[width=\hsize]{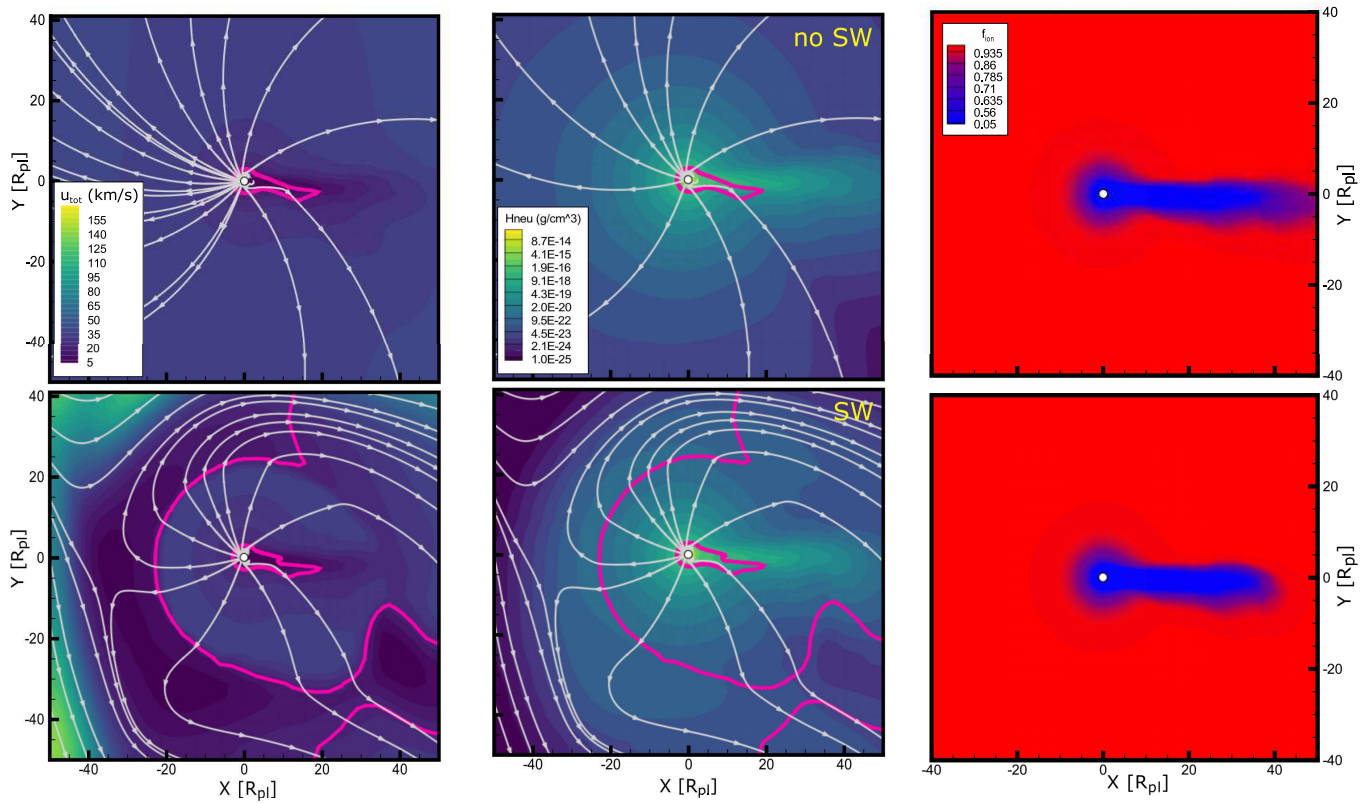}\\
  \caption{The 2D cut in the ecliptic plane of the model outputs for TOI-421~b. The top row shows the case of pure planetary outflow, and the bottom row shows the case including the stellar wind. The columns from left to right are total velocity, density of neutrals, and ionization fraction (zoomed in). The white streamlines show the atmospheric/stellar wind flows, and the magenta lines are the sonic surfaces.}\label{fig::toi421b}
\end{figure*}

\begin{figure*}
  \includegraphics[width=\hsize]{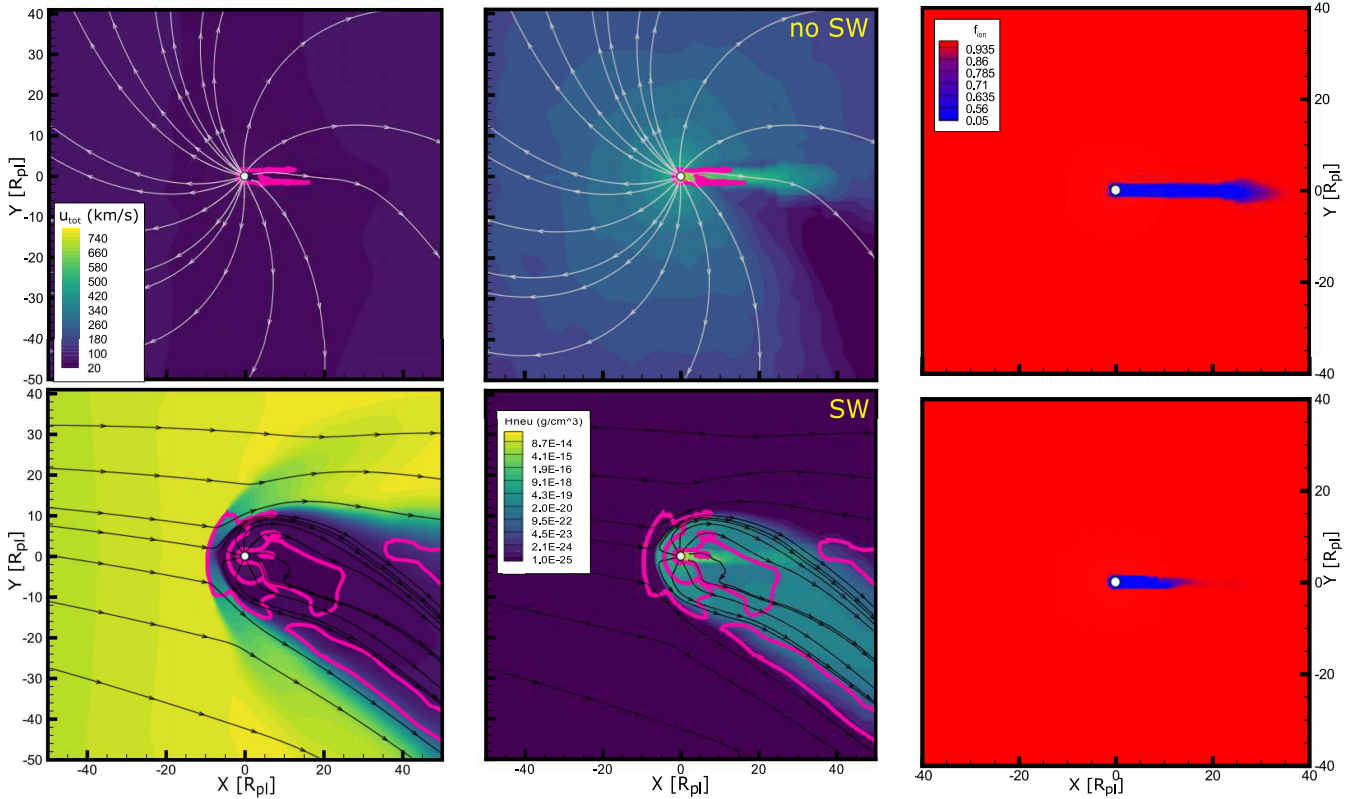}\\
  \caption{The same as in Figure~\ref{fig::toi421b}, but for TOI-942~b.}\label{fig::toi942b}
\end{figure*}

To model the escaping atmosphere of a planet, we start with modeling only the planetary outflow, including centrifugal and Coriolis forces. When the simulation reaches steady-state, we inject the stellar wind penetrating the simulation domain from the negative X direction. In Figures~\ref{fig::toi421b} and \ref{fig::toi942b}, we show the 2D (in ecliptic plane $Z=0$) distributions of total velocity, the density of neutral hydrogen, and the ion fraction at both of these stages for TOI-421~b and TOI-942~b, respectively. As the results for ``c'' planets in both systems are similar to those of ``b'' planets but less pronounced, we will focus mainly on ``b'' planets, when describing our results. The plots for ``c'' planets can be found in Appendix~\ref{apx:planets_c}. 

We should note that due to perturbations induced by the stellar wind, the simulation reaches only a quasi-steady-state: while the substellar region remains stable, the comet-like tail oscillates with a period of about 1000 steps {(for comparison, we note the typical convergence times of 4000-10000 steps, while we run the whole simulation until $\sim$20000 steps for planets in TOI-421 system and until $\sim$40000 steps for planets in TOI-942 system)}. These perturbations, however, do not affect the integral parameters of the atmosphere, such as atmospheric mass loss rate, which vary by less than $1\%$.

The streamlines in Figures~\ref{fig::toi421b} and \ref{fig::toi942b} depict the velocity vector projection onto the ecliptic plane, i.e., the atmospheric/stellar wind material flow, and the magenta line shows the position of the sonic surface, where $u_{\rm tot}$ equals the sound velocity. In the case of a pure planetary outflow, the atmospheric material escapes nearly radially, and the density distribution is mainly shaped by the planetary gravity and centrifugal and Coriolis forces. The flow starts subsonic from the planetary surface and accelerates to supersonic velocities before it reaches the Roche lobe (at $\sim 3$~$R_{\rm pl}$ for TOI-421~b and at $\sim 1.7$~$R_{\rm pl}$ for TOI-942~b), and the sonic surface stretches at the opposite side as far as $\sim 20$~$R_{\rm pl}$. With the inclusion of the stellar wind, the flows are deflected away from the star. The inner sonic surface remains almost unchanged (except for the case of TOI-942~b, opposite to the star), but a secondary sonic surface appears due to the interaction with the stellar wind. We discuss it in more detail in the next section.

\subsection{Interaction with stellar winds and the atmospheric mass loss}\label{ssec::results_sw}

Despite the similarity of the position and parameters of {the} planets in TOI-421 and TOI-942 systems, both planetary and stellar winds and their interaction look rather different due to the very different stellar ages (hence, different wind properties and stellar XUV {flux}). {The surface escape velocities of the inner planets are similar (likewise for the outer planets). This means that the gravitational acceleration within each pair of planets are similar, being the main difference in driving the planetary outflow the thermal gradients caused by the stellar XUV irradiation.}
In terms of the XUV insolation, young planets in the TOI-942 system are exposed to $\sim 20$ times higher flux than the older planets. This results in a consequently stronger outflow: we obtain the atmospheric mass loss rates (not accounting for stellar wind effects) of $5.05\times10^{11}$~${\rm g~s^{-1}}$ and $2.2\times10^{11}$~${\rm g~s^{-1}}$ for planets TOI-942~b and c, respectively, which are more than an order of magnitude higher than those of TOI-421~b and c, namely $3.2\times10^{10}$~${\rm g~s^{-1}}$ and $2.9\times10^{10}$~${\rm g~s^{-1}}$. At the same time, the outflow velocities remain similar. For the inner planets, it is about 28~${\rm km~s^{-1}}$ and 32~${\rm km~s^{-1}}$ at the L1 point for TOI-942~b and TOI-421~b, respectively. Thus, higher escape rates from younger planets are achieved due to the denser outflow (due to the higher stellar XUV/$L_{\rm bol}$), particularly in the tail. 

\begin{figure*}
  \includegraphics[width=0.8\hsize]{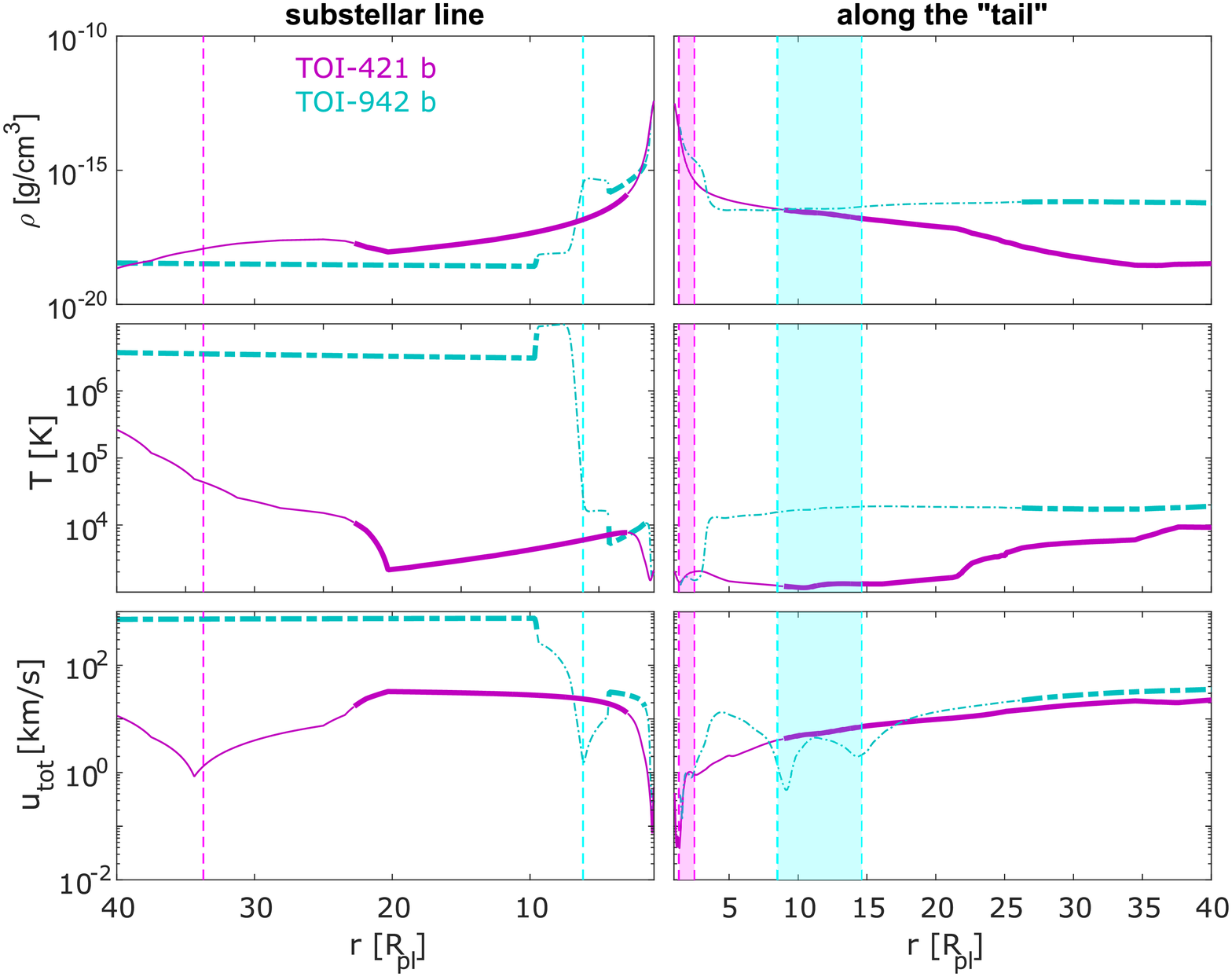}\\
  \caption{The 1D profiles extracted in the ecliptic plane for TOI-421~b (violet solid lines) and TOI-942~b (turquoise dashed-dotted lines) along the substellar line ($[X,Y]$ from $[-40,0]$ to $[-1,0]$) and along the ``tail'' of the escaping atmosphere ($[X,Y]$ from $[1,0]$ to $[49.5,-25.5]$/$[50, -3.8]$ for TOI-942~b/TOI-421~b, respectively). Planetary parameters are given against the radial distance, from top to bottom: the total density, temperature, and total velocity. Thin/thick lines represent the subsonic/supersonic regimes of the outflow. Dashed vertical cyan/magenta lines denote the positions of the sign reversal of radial velocity for TOI-942~b/TOI-421~b, respectively. The shaded rectangles in the right column show thus the small regions of reversed flows in the ``tail'' areas.}\label{fig::comp-421-942-b}
\end{figure*}

The wind of the young star TOI-942 is denser and hotter (as was discussed in Section~\ref{sec::target_systems}), and thus $\sim 8$ times faster at the orbit of the inner planet TOI-942~b than at TOI-421~b. This results in qualitatively different interactions between stellar and planetary winds. To ease the comparison, we show in Figure~\ref{fig::comp-421-942-b} the 1D profiles of total density, temperature, and velocity along the ``substellar'' line ($Y=0$ and $X<-1 R_{\rm pl}$) and along the comet-like tail of the outflow for both inner planets. The latter, corresponds to X ranging from 1 to 50~$R_{\rm pl}$ and Y ranging from -0.3 to 3.8~$R_{\rm pl}$ in case of TOI-421~b and $[X,Y]$ between $[1,0]$ and $[49.5, -29.5]$~$R_{\rm pl}$ in case of TOI-942~b.

In the case of the 9~Gyr old TOI-421~b, the interaction occurs between a strong planetary outflow and a relatively slow and weak stellar wind. The supersonic (from the source 1D wind model) stellar wind is not strong enough to balance the pressure of the planetary wind and the interaction occurs near the -X boundary. Thus, the stellar wind remains close to supersonic in the small region near the interaction (yellow regions in the bottom left panel of Figure~\ref{fig::toi421b}), but it is subsonic in the rest of the simulation domain.
The planetary wind decelerates to subsonic velocities as well at the distance of around $20 R_{\rm pl}$ along the substellar line (see bottom left panels in Figures~\ref{fig::toi421b} and \ref{fig::comp-421-942-b}), and {the} two flows meet at about $35 R_{\rm pl}$. The interaction region between both winds is stretched in space and starts at around $20 R_{\rm pl}$. As this point is above the planetary Roche lobe ($\sim10$~$R_{\rm pl}$ along the substellar line), the stellar wind does not affect the atmospheric mass loss rate (the same happens for planet c). The compression of the interacting winds increases the density of the flow between $\sim 20 R_{\rm pl}$, where the speed drops, and the point where two flows meet (Figure~\ref{fig::comp-421-942-b}, top left panel). {The temperature grows gradually along the substellar line from $\sim 20 R_{\rm pl}$ outwards and reaches the value of $\sim 0.3$~MK, which is lower than the stellar wind temperature of 1.3~MK due to the stretched interaction region and the strong planetary outflow moving it towards the left boundary.}

For the young planet TOI-942~b, the picture is very different. The dense and fast stellar wind compresses the atmosphere significantly, which is clearly seen in Figure~\ref{fig::toi942b}, and {the} two flows meet around the L1 point of the planet at $\sim 6 R_{\rm pl}$. The shock is well pronounced and localized within 4.2-10~$R_{\rm pl}$ in the direction towards the star. At the radial distance of 4.2-6~$R_{\rm pl}$, due to the compression, the velocity of the planetary outflow decreases steeply from 32 to about 1 ${\rm km~s^{-1}}$ (the flow becomes subsonic), the density of the flow increases from $1.5\times10^{-16}$~${\rm g~cm^{-3}}$ to $4.1\times10^{-16}$~${\rm g~cm^{-3}}$, and the temperature of the flow grows about 3 times from 5300 to 16000~K (see the left column of Figure~\ref{fig::comp-421-942-b}). From the side of the stellar wind, the same occurs: the speed of the wind drops from 750 to 1~${\rm km~s^{-1}}$, and thus the temperature of the shock increases to as high as $\sim 9$~MK. {Beyond 10~$R_{\rm pl}$, the modelling domain is fully filled with stellar material, as clearly seen in Figure~\ref{fig::comp-421-942-b} by almost constant $u_{\rm tot}$, $T$, and $\rho$ (in the ideal case of the isothermal wind the temperature would be constant, but in our case the adiabatic index is taken as 5/3 and it decreases by $\sim 10^5$~K between 10 and 40~$R_{\rm pl}$).}

This interaction between stellar and planetary winds pushes the planetary material towards the tail of the planetary outflow and thus creates a lot of perturbations, and {so} the sonic surfaces become complex. We can distinguish two dominant structures: the sonic surface surrounding the subsonic region around the planet (close to the planet surface in the stellar direction and stretched in the tail direction to about 20~$R_{\rm pl}$), and the subsonic region at the compression shock. These two structures remain rather stable, while the two ``flank'' subsonic regions in the tail part of the flow are affected by the perturbations in the tail and change their position with a period of $\sim 1000$ simulation steps, as we mentioned at the beginning of this section.

The atmospheric mass-loss rate from the planets in the young system TOI-942 changes with the inclusion of the stellar wind: it has a negligible growth from $5.05\times10^{11}$ to $5.1\times10^{11}$~${\rm g~s^{-1}}$ in the case of planet b, and grows from $2.2\times10^{11}$ to $3.2\times10^{11}$~${\rm g~s^{-1}}$ for planet c. This is likely due to the fast stellar wind pushing planetary material towards the tail and accelerating/dragging away the material on the flanks. However, these changes are insignificant in terms of planetary atmospheric evolution.

Due to the extreme XUV irradiation, the atmospheric material of younger planets is highly ionized. The ion fraction increases steeply with distance: if for TOI-421~b the atmosphere becomes fully ionized around the L1 point in the stellar direction, for TOI-942~b it occurs already at $\sim 2.5$~$R_{\rm pl}$. In Figures~\ref{fig::toi421b} and \ref{fig::toi942b}, one can see, that for the older planet the neutral material is present within 10~$R_{\rm pl}$ in Y, Z and negative X and up to $\sim 40$~$R_{\rm pl}$ in the positive X direction. For the young planet TOI-942~b, this distance decreases to $\sim 2$ and 25~$R_{\rm pl}$, respectively. 
Further on, the interaction between planetary and stellar winds creates a local zone of the planet-ward (reverse) radial velocity in the tail part of the planetary outflow at $\sim$9-15~$R_{\rm pl}$ (see the right column of Figure~\ref{fig::comp-421-942-b}, cyan shaded area). The change of the outflow geometry leads to the ionized material in the tail being pushed towards the planet as close as $\sim 10 R_{\rm pl}$ from the planet. 

This difference in ionization level has serious implications for observations of the atmospheric mass loss at such planets, as by Ly-$\alpha$ or H$\alpha$ absorption, one can detect only a neutral part of the atmosphere. Therefore, despite the atmospheric mass loss being an order of magnitude higher for younger planets, the transit signatures might look not too different in both cases. We discuss it, in more detail in Section~\ref{ssec::lya}.

\subsection{Decoupling the X-ray and EUV parts of the spectra}\label{ssec:results_xuv}

The common approach of modeling the stellar XUV heating used in this study is based on the assumption that the  XUV (X-ray+EUV) flux can be treated as the flux integrated over the whole spectra emitted at the single-photon energy (wavelength). Given that the high-energy part of the stellar spectra is in general unresolved, and the specific flux distribution among the different photon energies has a small effect on the atmospheric mass loss from typical Kepler planets \citep[see, e.g.,][]{guo2016,odert2020}, this approach appears to be reasonable.

However, there is something special about the high-energy spectrum of planets younger than a few hundreds of megayears, which can have implications for atmospheric heating and ionization. Most of the planets, observed and modeled so far are those orbiting stars older than 1-2~Gyr. For such stars, the EUV-luminosity is typically 5-10 times higher than  the X-ray luminosity. For young stars, on the contrary, the X-ray part of the flux is on the same level as the EUV or even dominates the XUV spectra \citep[see, e.g., ][ and the Figure~1 of this paper]{sanz2011,Johnstone2020mors}. In particular, for TOI-942, the stellar evolution models by \citet{Johnstone2020mors} predict the $L_{\rm X}\approx 1.6 L_{\rm EUV}$.

Why might this be important for atmospheric modeling? The photons of different energies are absorbed in the atmosphere at different heights, with higher energy photons (i.e., X-ray part of a spectrum) being absorbed deeper. Therefore, if accounting explicitly for the absorption of X-ray, the heating of the atmosphere starts closer to the planet. The overall relevance of this heating process depends on a few parameters: the relation between the EUV and X-ray part of the spectra, the photon absorption cross-sections for different photon energies and atmospheric species, and the wavelength-dependent heating efficiency.

For exactness, we will consider here the XUV insolation model similar to our basic model described in Section~\ref{sec::model}, but assuming that the XUV spectrum is split into ``X-ray'' and ``EUV'' intervals, and that the integrated flux in each interval is emitted at a single photon energy, 20~eV for the EUV part (as before) and 248~eV (i.e., 5~nm) for the X-ray part \citep[this approach was used in ][]{kubyshkina2018grid}. We thus have two heating sources, for which the sum of the two fluxes is exactly equal to the total flux in our basic model ($L_{\rm XUV} = L_{\rm EUV} + L_{\rm X}$)\footnote{The XUV fluxes in \citet{Carleo_toi421} and \citet{Carleo_toi942} were estimated from $L_{\rm X}$ luminosities using the relations given in \citet{linsky2014}. In the present study, we used the relations from more recent work by \citet{Johnstone2020mors}, that is based on a larger sample of stars, in particular at young ages.}. We will refer to this model from now on as the split-wavelength model.

For these photon energies, the cross-section of the absorption by an hydrogen atom of the X-ray part of the stellar flux is $\sim5\times10^{-4}$ {times} smaller than that for the EUV part of the spectra. For this case, the Equation~\ref{eq::heating} can be rewritten as
\begin{equation}\label{eq::heating_decoupled}
    {\cal H} = \sum_{i}\eta_{\rm i}F_{\rm i}e^{-\tau_{\rm i}}\sigma_{\nu_0,i}n_{\rm n}, \text{ where i = X, EUV}. 
\end{equation}

Thus, if all other parameters are identical, the heating in the X-ray range would be negligible in comparison to the heating by EUV, and such decoupling of X-ray and EUV parts of the stellar flux would only lead to the decrease of about a factor of two in the total heating rate. However, as the X-ray is absorbed deeper in the atmosphere, the density of the neutrals in Equation~\ref{eq::heating_decoupled} will be about an order of magnitude higher for X-ray than for EUV ($\sim 38$ times in case of TOI-942~b).  Additionally, the heating efficiency is wavelength-dependent \citep{dalgarno1999,MC2009,Shematovich2014}. Equation~\ref{eq::eta}, used in this study, predicts the heating efficiency of 0.32 for the EUV range, and about 0.945 for the X-ray. Taking all of the above into account, and if $L_{\rm X}$ of a star is high enough (i.e., in particular, for young close-in planets), considering a more realistic spectrum can have a considerable effect on the predictions of the hydrodynamic upper atmosphere models.

\begin{figure}
  \includegraphics[width=\hsize]{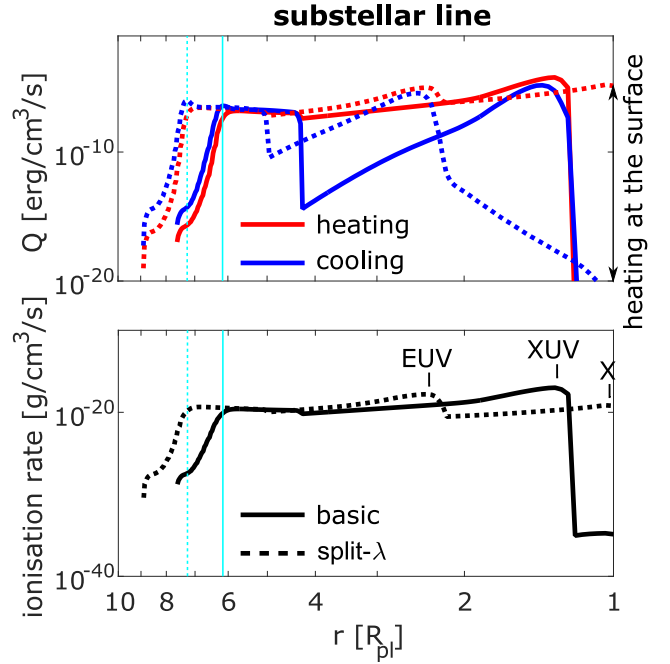}\\
  \caption{Comparison of the total heating/cooling volumetric rate profiles (top, red/blue lines, respectively) and ionization rate profiles (bottom) between the basic model (all XUV flux emitted at single energy of 20~eV, solid lines) and split-wavelength model (the stellar flux divided into X-ray part emitted at 248~eV and EUV part emitted at 20~eV, dotted lines) for TOI-942~b along the substellar line. The solid and dashed cyan vertical lines mark the position where the planetary wind meets the stellar wind for these two models, respectively.}\label{fig::XUV_split}
\end{figure}

In Figure~\ref{fig::XUV_split}, we compare the heating and cooling (top panel), and ionization rates (bottom panel) for TOI-942~b given by our basic model (solid lines) and the split-wavelength model (dotted lines). One can see, that the heating rate next to the planetary surface predicted by the split-wavelength model (see the red dotted line in the top panel) is comparable to the maximum heating rate given by the basic model (red solid line), which peaks at $\sim 1.3$~$R_{\rm pl}$. This leads to the increase of  temperature near the planet ($\sim$1.1-1.2~$R_{\rm pl}$) by about 1000~K in comparison to the basic model and to the consequent expansion of the atmosphere. This means that the density of the atmosphere (which is essentially the density of the neutral hydrogen, when  close to the planet) will be higher at the specific distance for the split-wavelength model than for the basic one. Therefore, the absorption depth of the EUV part of the spectra moves outwards, from $\sim 1.3$~$R_{\rm pl}$ in case of the basic model to $\sim 2.4$~$R_{\rm pl}$ in case of the split-wavelength model. The same can be seen also by the ionization rate profile (Figure~\ref{fig::XUV_split}, bottom panel, the basic XUV and split-wavelength EUV+X profiles).

The above has some effects on atmospheric parameters profiles shown in Figure~\ref{fig::comp-421-942-b}. Thus, the temperature maximum in the planetary outflow (see the peak on the right in the middle left panel of Figure~\ref{fig::comp-421-942-b}) shifts together with the maximum of the heating rate, from $1.55$~$R_{\rm pl}$ to $\sim 2.7$~$R_{\rm pl}$ with no significant change in the maximum temperature value ($\sim 3\%$ higher), and the shock between stellar and planetary wind (vertical dashed cyan lines in Figure~\ref{fig::comp-421-942-b}), and thus all the signatures described in Section~\ref{ssec::results_sw}, moves by about 1~$R_{\rm pl}$ outwards. The qualitative character of the atmospheric profiles remains, however, unchanged.

In terms of the atmospheric mass loss, the split-wavelength model predicts the atmospheric mass loss rate of $1.2\times10^{12}$~${\rm g~s^{-1}}$ from TOI-942~b, which is approximately two times higher than the prediction of the basic model. For the outer planet TOI-942~c, however, the increase in mass loss is much smaller and is only about 10\% higher than the escape rate in the basic model ($3.5\times10^{11}$~${\rm g~s^{-1}}$), due to the lower XUV fluxes (see Table~\ref{tab::systems}).

\begin{figure}
  \includegraphics[width=\hsize]{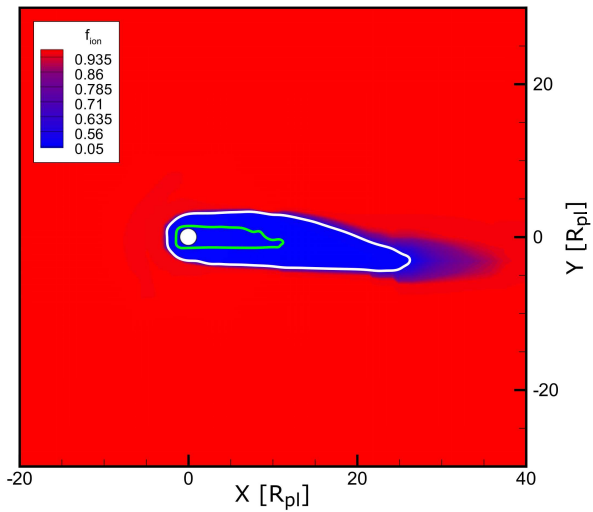}\\
  \caption{The 2D distribution of the ion fraction in the ecliptic plane for TOI-942~b assuming the split-wavelength X-ray+EUV model of the stellar irradiation. Designations are the same as in Figures~\ref{fig::toi421b} and \ref{fig::toi942b}. The white/green contour lines represent the $f_{\rm ion} = 0.5$ as given by the split-wavelength/basic models, respectively.}\label{fig::fion_decoupled_2d}
\end{figure}

However, the changes in the ionization fraction of the atmosphere are more important. The maximum in the ionization rate profile corresponding to the EUV part of the spectra shifts outwards (compare peaks marked as ``XUV'' for the basic and as ``EUV'' for the decoupled model in Figure~\ref{fig::XUV_split}). Thus, the atmosphere becomes fully ionized at a distance of about 3.5~$R_{\rm pl}$ instead of about 2~$R_{\rm pl}$ in the basic model. The ionization fraction close to the planet, before EUV ionization/heating becomes dominant, appears to be too low to change the atmosphere significantly: the ionization fraction below 2~$R_{\rm pl}$ remains lower than 1\%. This leads to the region dominated by neutral hydrogen (the blue-colored areas in the right columns of Figures~\ref{fig::toi421b} and \ref{fig::toi942b}, and in Figure~\ref{fig::fion_decoupled_2d}) being essentially more extended. Though in Figure~\ref{fig::XUV_split} we show the profiles along the substellar line, the effect is essentially 3D, and the area dominated by the neutral hydrogen increases approximately twice in all directions, as we show in Figure~\ref{fig::fion_decoupled_2d} for the ecliptic plane.

Thus, the inclusion of the (slightly) more realistic stellar spectra leads to the prediction of significantly higher Ly-$\alpha$ absorption, as we discuss in the next section. This is important, because it leads to larger chances of observing the atmospheric mass loss from the young planets in TOI-942 system.

\subsection{Synthetic Ly-$\alpha$ transits}\label{ssec::lya}

\subsubsection{Model}\label{ssec::lya-model}

To simulate the Ly-$\alpha$ absorption profiles for the planets considered in this study we employ the ray-tracing model developed in \citet{Vidotto2018,allan_vidotto2019} and applied in coupling with the variations of the 3D model used here in a series of works \citep{carolan2020,carolan2021}. The frequency-dependent optical depth along the line of sight (observer is placed in the positive X direction, opposite to the star) is given by

\begin{equation}
    \tau_{\nu} = n_{\rm n}\sigma_{\nu}\phi_{\nu}dX,
\end{equation}

\noindent where $\phi_{\nu}$ is the Voigt line profile function and $\sigma_{\nu} =  0.01103$~${\rm cm^2Hz}$ is the Ly-$\alpha$ absorption cross-section at the line center. Then, integrating over all rays transmitted through the atmosphere ($251\times251$ evenly spaced in Y and Z directions), we can obtain the frequency-dependent transit depth as 

\begin{equation}\label{eq::dF}
    dF_{\nu} = \int\int \frac{(1-e^{-\tau_{\nu}})}{\pi R_*^2}dYdZ.
\end{equation}

To quantify the absorption in the blue and red wings, we integrate the obtained $dF_{\nu}$ at the mid-transit over velocity in the intervals of [-300, -40] and [40, 300] km s$^{-1}$, respectively, excluding, therefore, the Ly-$\alpha$ line center dominated by the interstellar absorption and the geocoronal emission. 

\subsubsection{Absorption profiles}

\begin{figure}
  \includegraphics[width=\hsize]{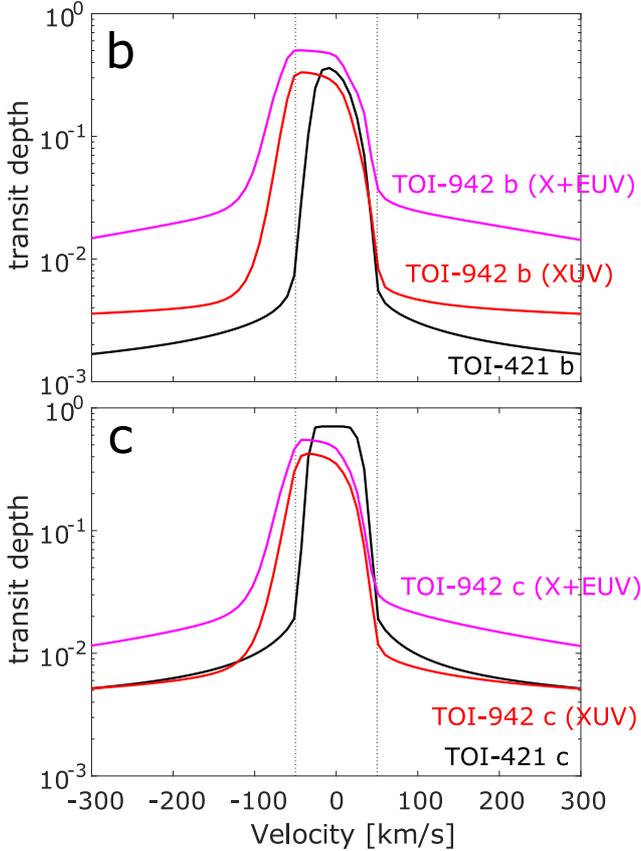}\\
  \caption{The theoretical Ly-$\alpha$ absorption profiles at mid-transit for b (top) and c (bottom) planets around TOI-421 (black lines) and TOI-942 (red and magenta lines for basic and split-wavelength models respectively). The black dotted vertical lines denote the [-40, 40]~${\rm km~s^{-1}}$ zone excluded from the estimations given in Table~\ref{tab::absorption}.}\label{fig::transit}
\end{figure}

In Figure~\ref{fig::transit}, we demonstrate the synthetic Ly-$\alpha$ calculated with Equation~\ref{eq::dF} for b (top panel) and c (bottom panel) planets in TOI-421 (black lines) and TOI-942 systems. For the latter, we consider both the results of basic (red lines) and split-wavelength (magenta lines) models, as discussed in Section~\ref{ssec:results_xuv}.

{To compare, the Ly-$\alpha$ absorption profiles for TOI-421 system were previously considered in \citet{Carleo_toi421} by employing the results of 1D hydrodynamic model by \citet{allan_vidotto2019}. For the same XUV fluxes as in the present paper, \citet{Carleo_toi421} predicted slightly higher escape rates of $4.5\times10^{10}$~${\rm g~s^{-1}}$ and $4.4\times10^{10}$~${\rm g~s^{-1}}$ for planets b and c, respectively, compared to  $3.2\times10^{10}$~${\rm g~s^{-1}}$ and $2.9\times10^{10}$~${\rm g~s^{-1}}$ from the present 3D study. However, the predicted maximum absorption for TOI-421~b is very similar in 1D and 3D cases: of 35\% and 36\%, respectively (see Figure~12 of \citealt{Carleo_toi421} and the top panel of Figure~\ref{fig::transit}). For TOI-421~c, though, the difference is significant. While the 1D model predicts the maximum absorption of 53\%, for the 3D model this value is $\sim$1.5 times larger, 71\%. This difference comes from the difference in the ion fractions predicted by the two models: as the absorption is only due to the neutral hydrogen, the maximum absorption is higher for lower ion fractions. As the 1D model considers only the star-planet line, where the irradiation is at maximum, it tends to overestimate the ionization of the atmosphere (and the mass loss rates, though to a lesser extent). This clearly demonstrates the importance of employing 3D modeling for this type of studies.}

When a planetary atmosphere interacts with a stellar wind, the neutral atmospheric material is pushed towards the observer (towards the tail part in Figures~\ref{fig::toi421b}, \ref{fig::toi942b}, and \ref{fig::fion_decoupled_2d}) and the Ly-$\alpha$ absorption profile becomes asymmetric and shifts towards the blue wing of absorption (negative velocity direction in Figure~\ref{fig::transit}). This shift is larger when the stellar wind is denser/stronger \citep[see, e.g.,][]{Carolina2021,carolan2021}. Thus, for both planets in the old system TOI-421, characterized by the relatively weak stellar wind and strong planetary outflows, the absorption profiles look nearly symmetric (which is easy to see by the vertical lines denoting $\pm 40$~${\rm km~s^{-1}}$ in Figure~\ref{fig::transit}). For planets in the young system TOI-942, the profiles are highly asymmetric (blue-shifted), both for the basic and split-wavelength XUV models, for which the profiles differ mainly quantitatively. In Table~\ref{tab::absorption} we show the integrated excess absorption ($\Delta F$, without the geometric transit) in the blue and the red wings, and calculate the total absorption as the sum of them.

\begin{table*}
\begin{tabular}{|c|c|c|c|c|}
  \hline
  planet & $\dot{M}$ [$10^{11}$~${\rm g~s^{-1}}$] & $\Delta F_{blue}$ [\%] & $\Delta F_{red}$ [\%] & $\Delta F_{tot}$ [\%] \\
  \hline
  TOI-421~b & 0.32            & 0.23  & 0.20 & 0.43 \\
  TOI-421~c & 0.29            & 0.63  & 0.63 & 1.26  \\
  TOI-942~b & 5.1             & 3.0   & 0.3  & 3.3  \\
  TOI-942~b (s-$\lambda$) & 12.0 & 7.8   & 2.0  & 9.8 \\
  TOI-942~c & 3.2             & 3.1   & 0.5  & 3.5 \\
  TOI-942~c (s-$\lambda$) & 3.5 & 6.1   & 1.5  & 7.6 \\
  \hline
\end{tabular}
\caption{The atmospheric mass loss rate, the excess absorption in the blue (-300 to -40 ${\rm km~s^{-1}}$) and red (40 to 300 ${\rm km~s^{-1}}$) wings, and the summed absorption for planets considered in this study. For planets in the TOI-942 system, we include the results for basic and split-wavelength XUV models (TOI-942~b (s-$\lambda$) and TOI-942~c (s-$\lambda$)).}
\label{tab::absorption}
\end{table*}

Another property of the strong stellar winds, discussed in detail in~\citet{carolan2021}, is to suppress the observable Ly-$\alpha$ signatures by confining the planetary outflow closer to the planet. This process is very effective for the TOI-942 system, as can be easily seen in Figure~\ref{fig::toi942b} (middle column). Therefore, despite the fact that assuming the basic XUV model, we obtain atmospheric escape rates of planets in the young system of $\sim 16$ and $\sim 11$ {times} higher than of planets in the old system (for planets b and c, respectively), the total absorption is only $\sim 8$ and $\sim 3$ times larger. Moreover, this difference is mainly produced by the blue-wing absorption, while the red-wing absorption is at a similar level for planets in both systems (as well as at the center of velocity profile).

Inclusion of the split-wavelength model for XUV heating, as we already discussed in Section~\ref{ssec:results_xuv}, leads to an increase in the atmospheric escape rates from planets in TOI-942 systems by $\sim 2$ and $\sim 1.1$ times compared to the basic XUV model for planets b and c, respectively. However, due to the significant changes in the ionization profiles, this leads to an even larger increase in the Ly-$\alpha$ absorption in the wings: by $\sim 3$ and $\sim 2.2$ times compared to the basic model. As the split-wavelength XUV insolation model appears to be more realistic for very young stars, it suggests that observing escape from such planets is of a higher probability than the basic model would have predicted.

This is important for the interpretation of observations. In particular, for TOI-942~c, two models predict significantly different absorption for the same stellar wind and nearly the same atmospheric mass loss. That implies, that, for young exoplanets, a more realistic assumption on the stellar high-energy spectra can be important if one wants to resolve the atmospheric mass loss from the observations.

\subsubsection{Perspectives for observations}

To further compare with the observations, we follow the same approach as \citet{hazra2021}. We consider the intrinsic Ly-$\alpha$ line profile of a star after absorption by the neutral ISM material ($F_{\rm out,\nu}$) as the out of transit line profile. This way, the flux density during the transit of a planet, including absorption by the ISM, can be written as
\begin{equation}
    F_{\rm in,\nu} = (1-dF_{\nu})F_{\rm out,\nu}.
\end{equation}
Finally, this value should be convolved with the response function of the specific instrument, to which we are interested to compare. We consider the STIS spectrograph on board the Hubble Space Telescope (HST) and employ the parameters for the convolution suggested in \citet{Bourrier2017}.

Since both stars lack observations in Ly-$\alpha$, we constructed the intrinsic profile for each star using the observationally-derived profile of other stars with similar spectral type, age and activity index. Usually these reference stars tend to be  closer than the TESS targets. However, we are assuming that the main ISM absorption will happen within a few parsecs, and so the effects from the ISM over the derived intrinsic profiles will be similar as the one found for the reference stars.

In the case of TOI-942, we scale by radius and distance the intrisinc Ly-$\alpha$ profile of the young K2~V dwarf star LQ Hya derived in \cite{wood2000}.
This is a fast rotating, $\sim 70$~Myr old star, located at  18.3~pc from the solar system, slightly smaller than TOI-942. Its radius is 0.69~$R_{\odot}$, and Ca II activity index ${\rm logR_{HK}}$ is $\sim-4.0$ \citep[][compare the latter to $\sim-4.17$ for TOI-942]{borosaikia2018}. The mass of the star is $0.8\pm0.1$~$M_{\odot}$ \citep{kovari2004,Tetzlaff2011}.
%
{To obtain the intrinsic profile of TOI-421 we used the same reference star as in \cite{Carleo_toi421}. Thus, we scaled the intrinsic profile of $\xi$ Boo A (G8 V, d = $6.70$ pc, $R_\star=0.78\,R_\odot$) derived by \citet{wood2005} to the radius and distance of TOI-421.}

\begin{figure*}
    \includegraphics[width=\hsize]{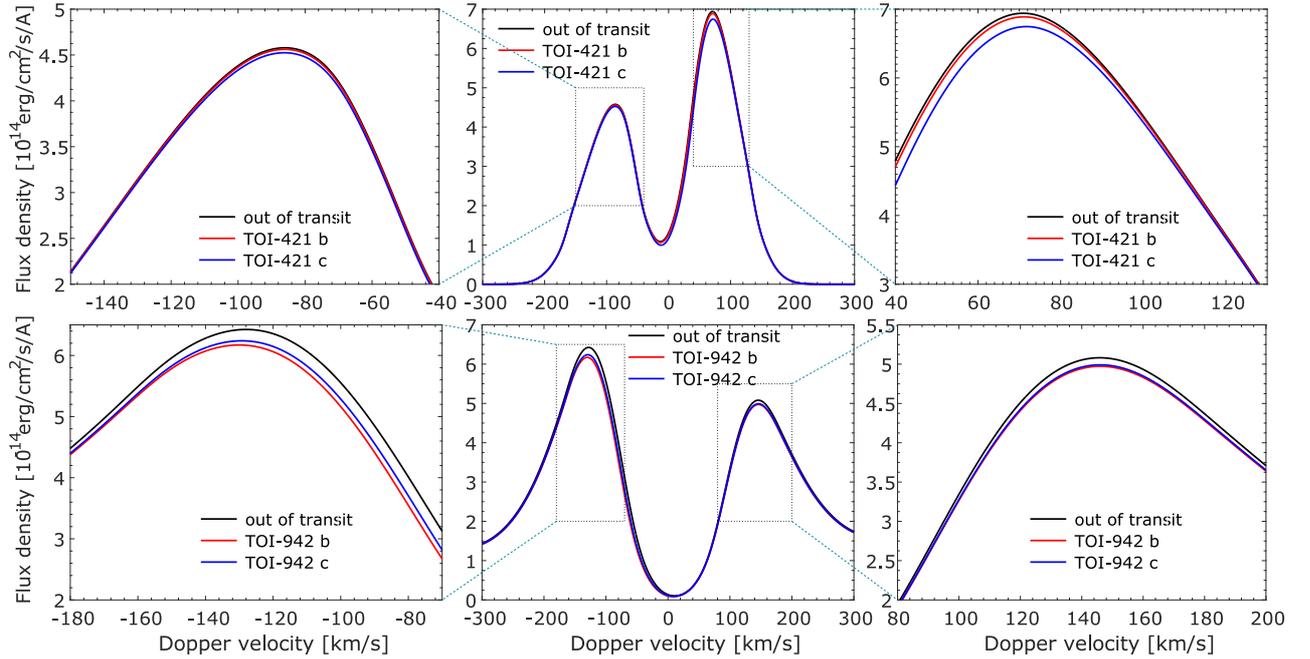} 
    \caption{Predicted Ly-$\alpha$ line profile at mid transit of TOI-421~b, c (top panels) and TOI-942~b, c (bottom panels) convolved with the instrument response function of STIS spectrograph on board of HST. The full profiles are shown in the middle column, and the zoom-in in blue and red wings (areas highlighted by the black rectangles in the middle column) is shown in the left and the right columns, respectively. The black line corresponds to the intrinsic stellar line profile, and red and blue lines show the flux densities during the transits of planets b and c of the systems, as shown in the legend.}
    \label{fig:lya-convolved}
\end{figure*}

In Figure~\ref{fig:lya-convolved}, we show the observational prediction of the flux density made for planets in TOI-421 (top panels) and TOI-942 (bottom) systems. In case of TOI-942 system, we consider the split-wavelength XUV model, as the more realistic one. The total decrease in the flux densities corresponds well to the predictions of the excess absorption in blue and red wings given in Table~\ref{tab::absorption}.
For both systems, the Ly-$\alpha$ line profiles look essentially asymmetric. Thus, in case of TOI-421 b and c, the peaks of Ly-$\alpha$ profiles are located at about +70 and -90 ${\rm km~s^{-1}}$, and the absorption is predominantly red-shifted. The planets in TOI-942 system look more appealing for observation, as the blue-wing absorption is dominating. The peaks in Ly-$\alpha$ profiles are shifted, however, to larger Doppler velocities (about +150 and -130 ${\rm km~s^{-1}}$).

%
Another issue that might complicate the observations is the high variability of young stars, which prevented more accurate mass constraints of the planets in TOI-942 system being made. The high flaring rate, typical for young stars, can significantly affect the observations \citep[see][]{hazra2021}.

\section{Discussions and conclusions}\label{sec::conclusions}
In the present paper, we have performed 3D modeling of the escaping atmospheres and their interaction with the stellar wind for two planets in the very young system TOI-942 and for two planets in the old system TOI-421. Despite the large similarity between the two systems (the planets are at nearly equivalent orbits and in a similar mass range, and the stars have similar masses, though rather different XUV luminosities typical for their ages), both the atmospheric mass loss and the interaction with stellar winds, are very different.

{To date, there is a number of studies dedicated to the 3D interaction between planetary outflows and stellar winds \citep{bisikalo2013,Schneiter2016,Carroll-Nellenback2017,Carolina2018,Carolina2021,McCann2019,Khodachenko2019,Esquivel2019,Debrecht2020,carolan2021,carolan2021r}. These models can be grouped into local planet-centered models where the stellar wind is set as an external boundary condition (as ours) and global models that include both the star and the planet in the computation domain \citep[see][]{Carroll-Nellenback2017, Carolina2018,Carolina2021,Esquivel2019}. {The benefit of the latter approach is that it can account for  effects non-accessible in the planet-centered frameworks, such as the formation of a disk around the star \citep{Debrecht2018}.} However, the former approach allows for better resolution close to the planet and {their smaller simulation domain usually implies that more complex physical processes, such as self consistently including heating and cooling processes, and ionisation balance, can be included in the simulation without being too computationally expensive}.}

{All the models above predict similar general signatures of the escaping atmospheres, including the asymmetry due to orbital motion, the tail-like structure, the photo-ionization shadow opposite to the star, and the blue-shifted absorption profiles shaped by the neutral hydrogen in the planetary outflow. Most of them consider hot Jupiters and assume the typical conditions of evolved main-sequence stars. Some, however, consider different stellar wind conditions. Therefore our models for the young planetary system TOI-942 are similar to the strong wind cases considered in \citet{Schneiter2016}, \citet{McCann2019}, and \citet{carolan2021}, with strongly confined planetary atmospheres and pronounced hydrodynamic shocks. Our models for the older system TOI-421 are similar to the weak stellar wind cases in the studies cited above, and the cases considered in \citet{Esquivel2019}.}

The escape rates from the planets in TOI-421 system are $3.2\times10^{10}$ g s$^{-1}$ and $2.9\times10^{10}$ g s$^{-1}$. In this situation, the strong planetary winds interact with a relatively weak stellar wind (comparable to the present-day solar wind). The interaction occurs far above the planets, and therefore has a minor effect on the upper atmospheres of the planet and no considerable effect on the planetary atmospheric mass loss. The stellar wind pushes the planetary outflow streams away from the star, but it does not lead to the noticeable asymmetries/changes in the theoretically predicted Ly-$\alpha$ absorption profiles compared to the no-wind case. 
{In terms of observations, for TOI-421 system, we predict potentially observable absorption in the Ly-$\alpha$ line, in particular for planet c. Despite the atmospheric mass-loss rates predicted by our models being similar for both planets, the atmosphere of planet b is significantly more ionized than the atmosphere of the outer planet, and the Ly-$\alpha$ absorption only occurs for the neutral part of the atmosphere. This leads to the difference in excess absorption of $\sim$3 times between planets b and c: $\sim0.4$\% and $\sim1.3$\%, respectively. This value is relatively small compared to Ly-$\alpha$ absorption observations of some hot Jupiters (e.g., HD~189733~b, 55~Cancri~b, HD~209458~b) and warm Neptunes (e.g., GJ~3470~b, GJ~436~b). 
{For HD~209458~b, \citet{VM2003} and \citet{VM2004} observed Ly-$\alpha$ absorption of about 5 to 6\%. 
For HD~189733~b, \citet{Lcv-d-Etangs2010} and \citet{Lcv-d-Etangs2012} obtained for different epochs  absorptions of $5\%$ and $3\%$, and later \citet{bourrier2013} reported excess absorptions of $2\%$ and  $5\%$ for the observations in 2010 and 2011, respectively.} For the hot Jupiter 55~Cancri~b, the reported absorption was $7.5\%$ \citep{Ehrenreich2012}. For the warm Neptune GJ~436~b, 
\citet{Ehrenreich2015} reported an absorption of $56\%$ and, finally, recent studies of GJ~3470~b revealed  Ly-$\alpha$ absorptions of $35\%$ \citep{bourrier2018,bourrier2021}. In all these cases, the observed absorption is considerably higher than the prediction made in this study for TOI-421~c and varies considerably for different observation epochs (i.e., current stellar activity levels). With this in mind, we expect that the excess absorption of $\sim1.3$~\% predicted for the quiet stellar conditions does not exclude the possibility of Ly-$\alpha$ absorption detection for TOI-421~c. In addition, we predict rather high values of absorption at the line centre ($\sim70\%$), but this part of the line is not observed due to the ISM absorption and geocoronal emission.}

Atmospheric mass-loss rates in the young system TOI-942 are about {an} order of magnitude higher than those in the old system, which is similar to the relation of the XUV fluxes of the two stars. Without the stellar wind, the mass loss rates are $4.5\times10^{11}$ g s$^{-1}$ and $2.2\times10^{11}$ g s$^{-1}$ from planets b and c, respectively. In turn, the interaction with the stellar wind occurs differently compared to the old system. The dense and fast wind from the young star confines the planetary atmosphere and creates a well-pronounced shock wave in the substellar region. The interaction occurs close to the planet: the boundary between stellar and planetary winds is located at approximately the L1 point in the case of the inner planet TOI-942~b, and within the Roche lobe in the case of the outer planet TOI-942~c. This leads to the erosion of the planetary upper atmosphere, and the corresponding increase in the atmospheric mass loss rates by $\sim22$~\% and $\sim45$~\% for planets b and c, respectively. Further on, the confinement of the atmospheres by the stellar wind, together with a higher ion fraction, leads to the suppression of the possible observable Ly-$\alpha$ signatures. Thus, despite the much higher atmospheric mass-loss rates, the predicted Ly-$\alpha$ excess absorption for the standard model (where the integrated XUV flux is emitted at the single wavelength) is only a few times higher than in the TOI-421 system: 3.3~\% and 3.5~\% for TOI-942~b and c, respectively.

{The above estimates, however, do not take into account that the high-energy (XUV) spectra of young stars are different in comparison to the spectra of  more evolved main-sequence stars. 
Though a detailed spectrum is in general not available, it is known that in {the} spectra of young stars the ratio between the X-ray and the EUV part is essentially higher than for old stars. According to various models and observations \citep[see, e.g.,][]{sanz2011,linsky2014,Johnstone2020mors}, for young stars, the energy in the X-ray wavelength range can be as much as 1-3 times of that in the EUV range, while for stars of a few Gyr, $L_{\rm X}$ is only a small fraction of XUV, typically up to an order of magnitude lower than $L_{\rm EUV}$.
This {turns out to be} important for modeling and the prediction/interpretation of the observations.}

We tested this possibility by implementing the split-wavelength XUV model, as described in Section~\ref{ssec:results_xuv}. 
We find, that applying the more realistic split-wavelength model leads to higher escape rates: about twice in the case of the inner planet TOI-942~b, and by about 6\% in the case of the outer planet TOI-942~c. 
The effect on the predicted Ly-$\alpha$ absorption is even higher, because of the atmospheres being more expanded due to the effective heating closer to the planetary surface and less ionized, as the main ionization processes occur further away from the planet. 
{Thus, the theoretical prediction of the total absorption increases about 3 times for planet b (from 3.3 to 9.8\%), and more than twice for planet c (from 3.5 to 7.6\%).} 

{This is consistent with the conclusions made by \citet{guo2016}, who tested the influence of the spectral energy distribution on the upper atmosphere of the hot Jupiter HD~189733~b using the 1D hydrodynamic models assuming hydrogen-helium atmosphere. They found that the mass loss from the planet is mainly determined by the total energy in the XUV range, similar to the more moderate case of TOI-942~c considered here. At the same time, the ionization profile is strongly affected by the shape of the spectrum, with the transition region between H and ${\rm H^+}$ moving outwards for spectra dominated by low-energy radiation, which can explain the differences in observations for different epochs. The activity level of the host star HD~189733 suggests an age of 1-2~Gyr. This demonstrates, that the usage of more realistic spectra can have implications also for more evolved planetary systems, though the most substantial effect is yet expected for the young ones, for which well-resolved spectra are normally not available. The typical targets for Ly-$\alpha$ observations so far were a few Gyr old planets (thus, most of the planets cited earlier are in the range of 1-5~Gyr old). With the ongoing improvement of instrumentation, however, we expect the fraction of the young planets among new detection and follow-up observation (for more detailed characterisation) to increase rapidly.}

\section*{Acknowledgements}
This project has received funding from the European Research Council (ERC) under the European Union's Horizon 2020 research and innovation programme (grant agreement No 817540, ASTROFLOW). The authors wish to acknowledge the SFI/HEA Irish Centre for High-End Computing (ICHEC) for the provision of computational facilities and support. This work used the BATS-R-US tools developed at the University of Michigan Center for Space Environment Modeling and made available through the NASA Community Coordinated Modeling Center.

\section*{Data Availability}
The data underlying this article will be shared on reasonable
request to the corresponding author.







\appendix
\section{Plots for c planets}\label{apx:planets_c}
%

\begin{figure*}
  \includegraphics[width=0.92\hsize]{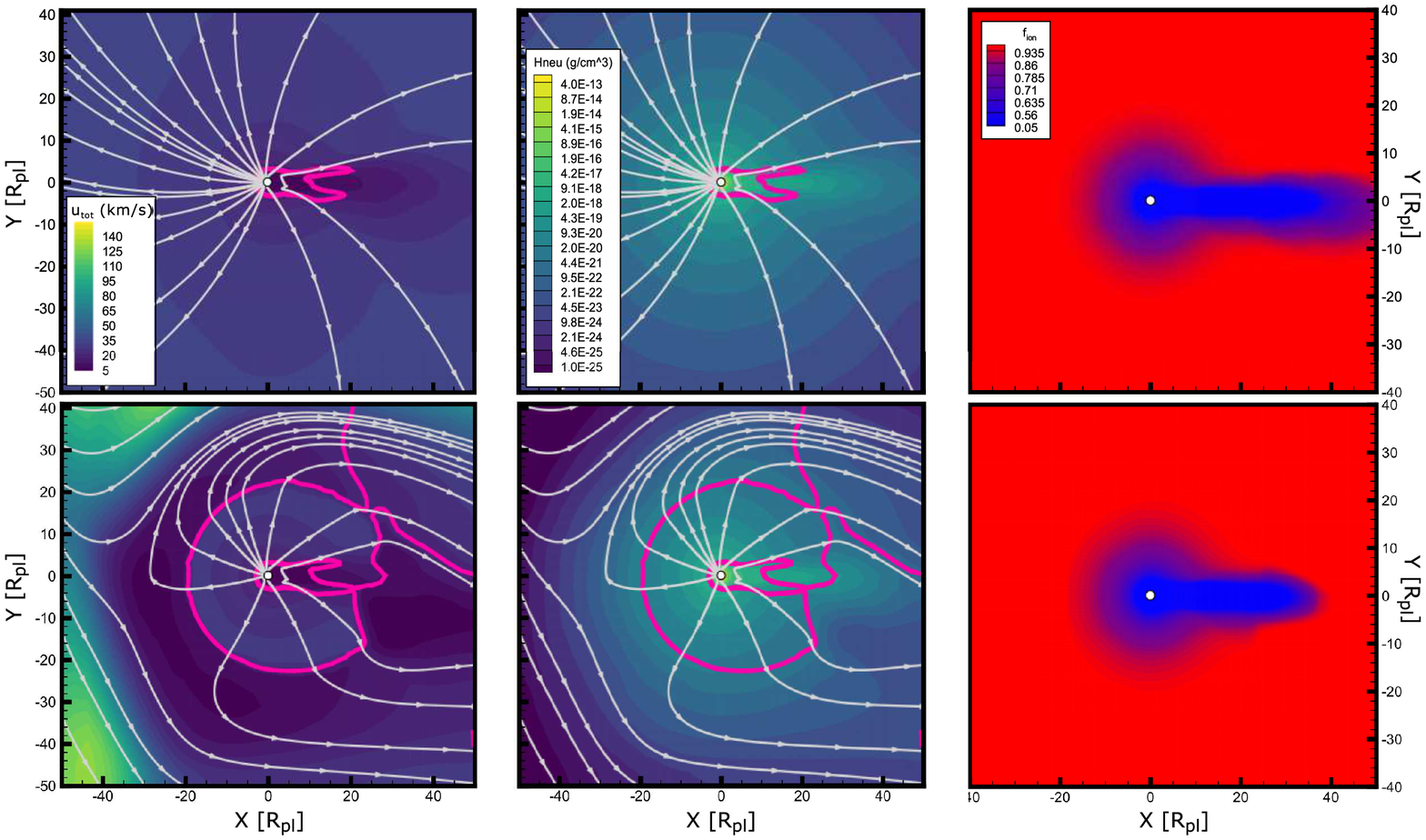}\\
  \caption{Same as Figure~\ref{fig::toi421b}, but for TOI-421~c.}\label{fig::toi421c}
\end{figure*}

\begin{figure*}
  \includegraphics[width=0.92\hsize]{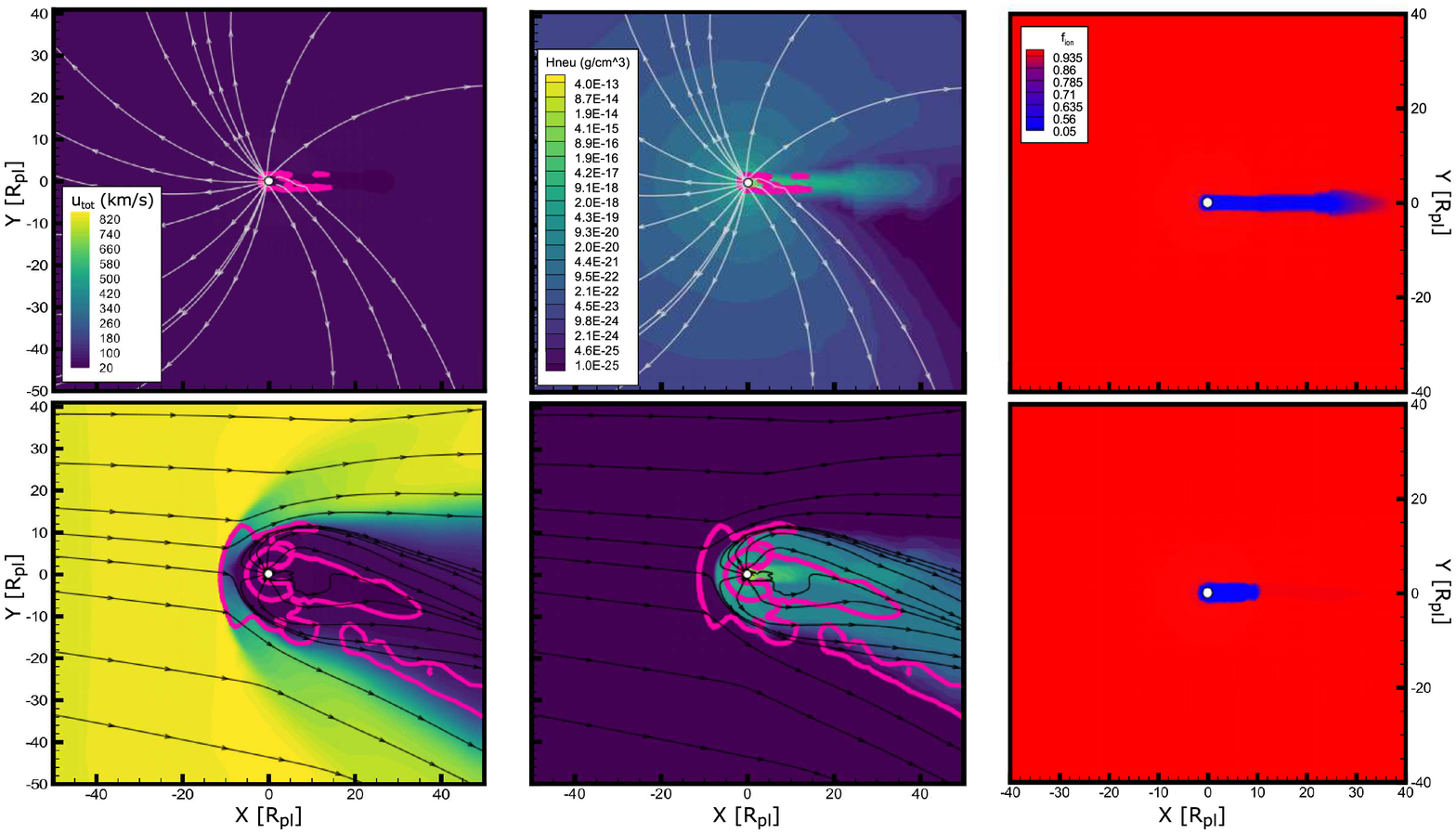}\\
  \caption{Same as Figure~\ref{fig::toi421b}, but for TOI-942~c.}\label{fig::toi942c}
\end{figure*}

\begin{figure*}
  \includegraphics[width=0.8\hsize]{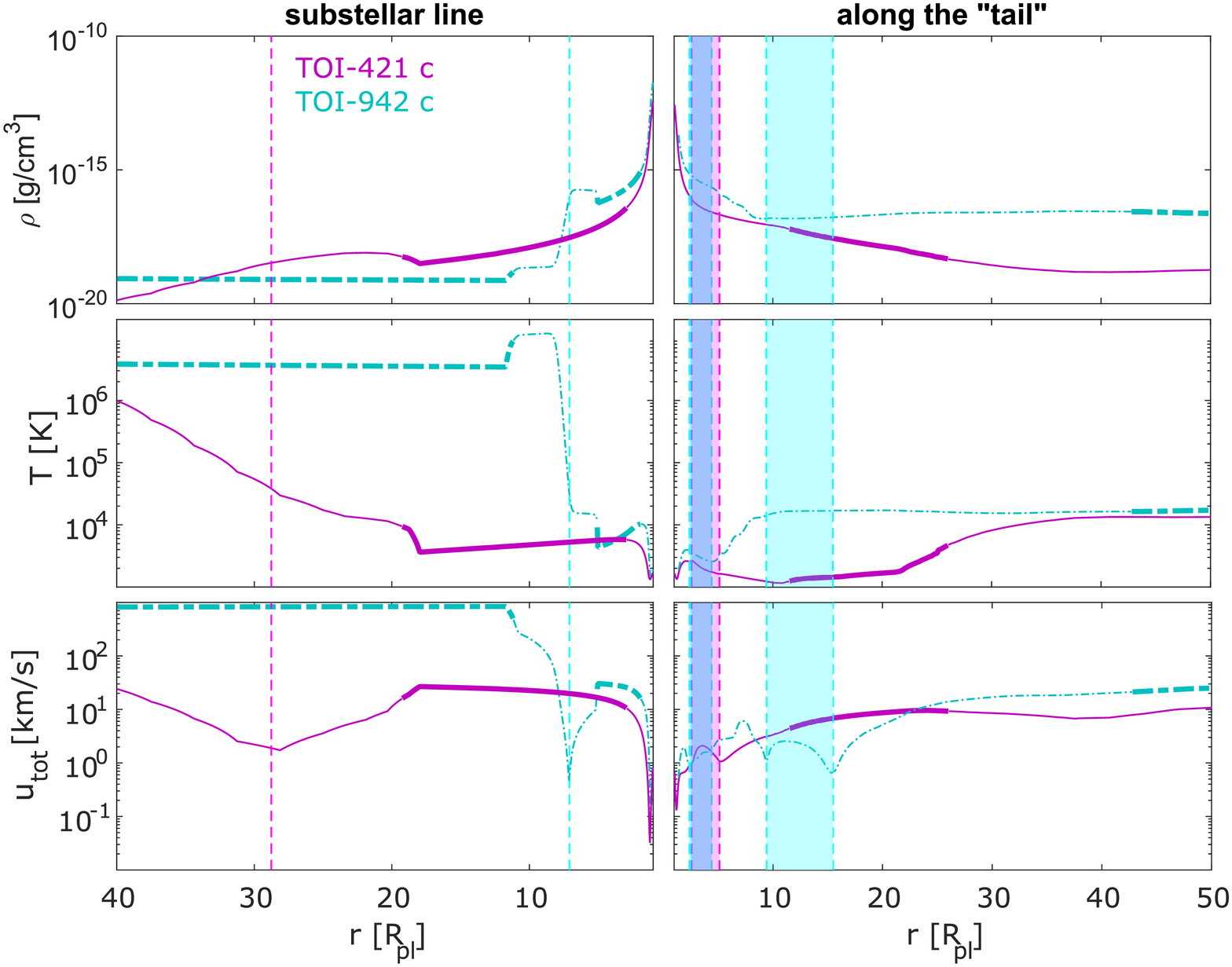}\\
  \caption{Same as Figure~\ref{fig::comp-421-942-b}, but for planets TOI-421~c (solid lines) and TOI-942~c (dashed-dotted lines).}\label{fig::comp-421-942-c}
\end{figure*}

Here, we present the plots equivalent to the ones considered in Section~\ref{sec::results}, but for c planets of TOI-421 and TOI-942 systems. Thus, in Figure~\ref{fig::toi421c}, we present the 2D projection in the ecliptic plane of the total velocity, the neutral hydrogen density, and the ion fraction for TOI-421~c. In Figure~\ref{fig::toi942c} we show the same for TOI-942~c. Finally, in Figure~\ref{fig::comp-421-942-c} we demonstrate the comparison equivalent to the one shown in Figure~\ref{fig::comp-421-942-b}, but for the c planets of {the} two systems. We point, that in the case of TOI-942~c, the interaction with the stellar wind occurs within the Roche lobe of the planet (L1 point is at the distance of about 11.6 planetary radii from the planet).

\bsp    
\label{lastpage}
\end{document}